\documentclass[journal, twoside, final]{IEEEtran}
\usepackage{cite}
\usepackage{graphicx}
\usepackage{amssymb, amsmath, amsthm}
\usepackage{amsfonts}
\usepackage{multirow}
\usepackage{mathrsfs}
\usepackage{color}
\usepackage{url}
\usepackage{flushend}
\usepackage{setspace}
\usepackage{placeins}
\usepackage{booktabs, multicol, multirow}
\allowdisplaybreaks

\newtheorem{lemma}{Lemma}
\newtheorem{theorem}{Theorem}
\newtheorem{remark}{Remark}
\newtheorem{corollary}{Corollary}

\begin{document}

\title{On the Efficiency of \\ Far-Field Wireless Power Transfer}

\author{Minghua~Xia,~\IEEEmembership{Member,~IEEE}, and Sonia~A\"{\i}ssa,~\IEEEmembership{Senior Member,~IEEE}
\thanks{%
Manuscript received July 10, 2014; revised January 17, 2015; accepted March 09, 2015. Date of publication March 27, 2015. The associate editor coordinating the review of this manuscript and approving it for publication was Dr. Akbar Sayeed. This work was supported by a Discovery Grant from the Natural Sciences and Engineering Research Council (NSERC) of Canada. This work was done while M. Xia was with the Institut National de la Recherche Scientifique (INRS), University of Quebec. Part of this work was presented at the IEEE Global Communications Conference (Globecom), Austin, TX, USA, December 7Ð11, 2014.}
\thanks{%
M. Xia was with the Institut National de la Recherche Scientifique (INRS), University of Quebec, Montreal, QC H5A 1K6, Canada; he is now with Sun Yat-sen University, Guangzhou, 510275, China (e-mail: xiamingh@mail.sysu.edu.cn).
}
\thanks{%
S. A\"issa is with the Institut National de la Recherche Scientifique (INRS), University of Quebec, Montreal, QC H5A 1K6, Canada (e-mail: aissa@emt.inrs.ca).
}
\thanks{%
Color versions of one or more of the figures in this paper are available online at http://ieeexplore.ieee.org.
}
\thanks{%
Digital Object Identifier 10.1109/TSP.2015.2417497}
}

\markboth{IEEE Transactions on Signal Processing, accepted for publication} {Xia \MakeLowercase{\textit{et al.}}: On the Efficiency of Far-Field Wireless Power Transfer}

\maketitle

\pubid{1053-587X~\copyright~2015 IEEE. Personal use is permitted, but republication/redistribution requires IEEE permission.}

\pubidadjcol

\begin{abstract}
\noindent  Far-field wireless power transfer (WPT) is a promising technique to resolve the painstaking power-charging problem inherent in various wireless terminals. This paper investigates the power transfer efficiency of the WPT segment in future communication systems in support of simultaneous power and data transfer, by means of analytically computing the time-average output direct current (DC) power at user equipments (UEs). In order to investigate the effect of channel variety among UEs on the average output DC power, different policies for the scheduling of the power transfer among the users are implemented and compared in two scenarios: homogeneous, whereby users are symmetric and experience  similar path loss, and heterogeneous, whereby users are asymmetric and exhibit different path losses. Specifically, if opportunistic scheduling is performed among $N$ symmetric/asymmetric UEs, the power scaling laws are attained by using extreme value theory, and reveal that the gain in power transfer efficiency is $\ln{N}$ if UEs are symmetric whereas the gain is $N$ if UEs are asymmetric, compared with that of conventional round-robin scheduling. Thus, the channel variety among UEs inherent to the wireless environment can be exploited by opportunistic scheduling to significantly improve the power transfer efficiency when designing future wireless communication systems in support of simultaneous power and data transfer.
\end{abstract}


\section{Introduction}
\label{Section:Introduction}
\subsection{Context and Motivation}
\IEEEPARstart{C}{ompared} to the rapid development of wireless data transfer technology, the power transfer technique in support of wireless communication systems remained stagnant in the past decades. Indeed, wired charging remains the main way to feed a device battery, which is a painstaking routine for mobile users and can even be a challenge, especially when users are on the move or cannot afford interruption in their wireless data service. In particular, when the battery is used out and cannot get recharged on time, an abrupt service interruption does not only degrade the quality of service dramatically, but can also yield a loss in users' data which is unaffordable in critical applications. As a result, the shortage of battery endurance constitutes a major bottleneck which hinders the development of ubiquitous wireless communication systems with access to services whenever and wherever needed.

In order to resolve the power-charging problem inherent in current wireless terminals, it is necessary to equip them with far-field wireless power charging capability. It is not hard to imagine that, if, for example, a mobile phone or a laptop can be remotely charged whenever it accesses a wireless network, mobile users will never worry about power shortage. Clearly, the capability of far-field wireless power charging will be infinitely attractive to mobile users and will trigger a revolution in the way wireless communication systems of the future will function and operate.

Current techniques of wireless power charging, which are mainly based on the principle of electromagnetic inductive coupling, are essentially {\it near-field} techniques.  For instance, in the {\it Qi} standard, the distance between a charging device, e.g. a mobile phone, and its power transfer pad should not exceed 4 cm (1.6 in) \cite{Qi}. Also, the power transfer pad must be connected to an electrical outlet in a wired way. Obviously, compared to traditional wired charging, the near-field wireless power charging can reduce neither the users' concern on the shortage of battery endurance nor the users' labour in the charging.

\pubidadjcol

To extend the distance of wireless power transfer (WPT), magnetic resonance and/or radio frequency (RF) based {\it far-field} power transfer are promising. The concept dates back to the late 1800s. Indeed, the earliest prototype of magnetic resonance based WPT was demonstrated by Heinrich Hertz in 1887 \cite[Fig. 24]{Hertz1962}. In practice, Nikola Tesla attempted to wirelessly transmit power by magnetic resonance at Colorado Springs, CO, USA, in 1899. In this experiment, however, no evidence was collected on whether any significant amount of power would be available at any distant point. The first successful RF based power transfer was performed by Harrell V. Noble at the Westinghouse Laboratory, which was demonstrated to the general public at the Chicago World's Fair of 1933--1934 \cite{Brown84TMTT09}. Later on, RF based far-field WPT technology found extensive applications where no human interaction is needed, for instance, aeronautics and aerospace. Among them, the most famous experiment was performed by William C. Brown in 1963, who wirelessly powered a helicopter flying at 60 feet above the ground level. Another ambitious WPT experiment was the Stationary High Altitude Relay Platform (SHARP), which was performed by Communications Research Centre of Canada in 1987. In this experiment, RF based WPT was used to provide energy from ground station to SHARP at about 70,000 feet altitude. Then it was not until 2008 that simultaneous wireless transport of power and data in communication systems was proposed, by Lav R. Varshney at MIT \cite{VarshenyISIT08}. Although magnetic resonance and/or RF based WPT techniques have already found many applications in practice \cite{Shoki13PIEEE06} \cite[Chap. 4]{Shinohara14}, potential health risks of human exposure to electromagnetic fields constitute a major factor which needs further R\&D efforts to achieve the full potential of these techniques and a widespread commercial deployment. This safety issue is beyond the scope of this paper and the interested reader is referred to the up-to-date survey in \cite{Christ13PIEEE06}. Since magnetic resonance based WPT and RF based WPT have different design methodologies, in the rest of the paper the term WPT is limited to the latter, i.e. WPT via radiowaves, unless otherwise stated.

\begin{figure}[t]
\centering
\includegraphics [width=3.5in, clip, keepaspectratio]{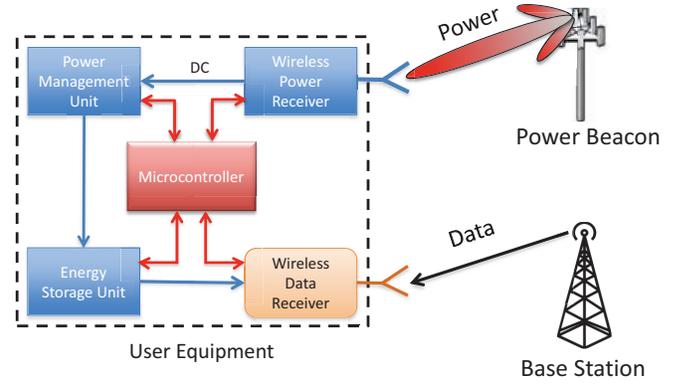}
\caption{A wireless system in support of simultaneous power and data transfer.}
\label{Fig_ReceiverModel}
\end{figure}

Figure~\ref{Fig_ReceiverModel} shows a wireless system in support of simultaneous power and data transfer, where the block diagram of a user equipment (UE) consisting of a specific power receiver and a traditional data receiver is sketched. In this system, a conventional base station (BS) exchanges data with the UE while a newly deployed power beacon radiates power to the UE. In practice, power beacons can be deployed separately from BSs or be integrated into BSs. In particular, power beacons can be integrated into femtocell BSs, since they both serve nearby users within the area of radius on the order of ten meters. With such a dedicated power beacon, power charging cables are not compulsory anymore and UEs can be remotely charged without the users' intervention.  Although this system is promising for next-generation wireless communications, there are few related papers in the open literature with most of them dedicated to calculating the achievable power versus data rate region, from an information theory perspective. Specifically, the original work \cite{VarshenyISIT08} focused on non-fading scenario, which was extended to frequency-selective fading by considering a coupled-inductor circuit problem in \cite{GroverISIT10}. Lately, the authors of \cite{Zhou13TWC11} studied how to design power/data receivers in a separate or an integrated way. Based on the separate structure in \cite{Zhou13TWC11}, optimal switching between the power receiver and the data receiver was addressed in \cite{Liu13TWC01}, to achieve various tradeoffs between power transfer and data reception.

In practice, in order to avoid interference between the power receiver and the data receiver of a UE, these units should work in different frequency bands. Actually, since far-field WPT is still in its infancy, so far no particular frequency band has been specified yet for this technique. Thus, so far the industrial-scientific-medial (ISM) radio bands remain the only option in prototype development. For example, in \cite{Falkenstein11TCS12}, 5.8 GHz and 2.45 GHz were used for power transfer and data transmission, respectively. Accordingly, at the UE side, the power receiver and the data receiver are separately designed, as shown in Fig.~\ref{Fig_ReceiverModel}. Generally speaking, systems in support of simultaneous power and data transfer at different frequency bands are known as {\it out-band systems} whereas those working at the same frequencies are known as {\it in-band systems}. Clearly, out-band systems are free of interference between the power and data transfer segments.

\subsection{Efficiency of Far-Field Wireless Power Transfer}
\begin{figure}[t]
\centering
\includegraphics [width=2.5in, clip, keepaspectratio]{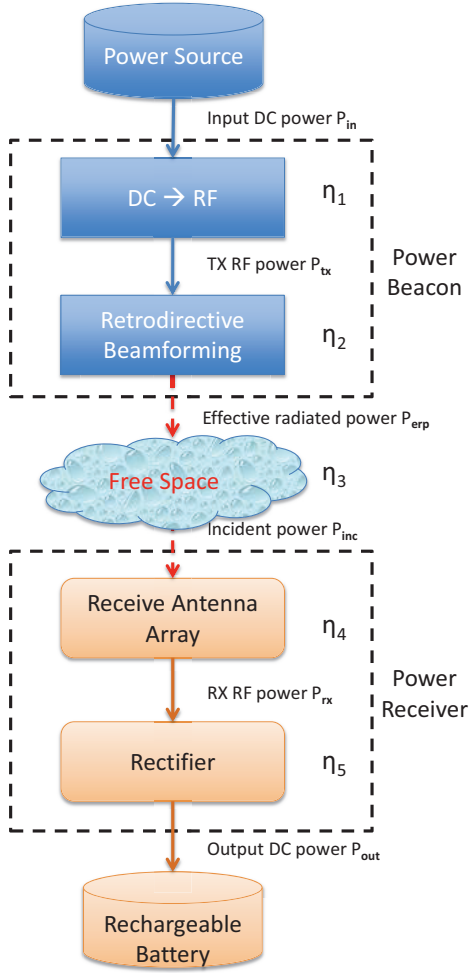}
\caption{Block diagram of a WPT segment in an out-band system.}
\label{Fig_WPT}
\end{figure}

Figure~\ref{Fig_WPT} shows the inner structure of a WPT segment in an out-band system (corresponding to the link from the power beacon to the power receiver shown in Fig.~\ref{Fig_ReceiverModel}), which consists of five subsystems: (1) DC-to-RF conversion, (2) retrodirective beamforming at the transmit antenna array, (3) free-space transmission, (4) receive antenna array and (5) rectifier. Thus, the end-to-end power transfer efficiency, $\eta_\mathrm{e2e}$, can be expressed as the product of $\eta_i$, $\forall i \in [1, 5]$, where $\eta_i$ denotes the power transfer efficiency at the $i^{\mathrm{th}}$ subsystem. That is,
\begin{equation}
\label{Eq.e2eEfficiency}
\eta_\mathrm{e2e} \triangleq \frac{P_\mathrm{out}}{P_\mathrm{in}} = \prod_{i=1}^5{\eta_i},
\end{equation}
where $P_\mathrm{in}$ and $P_\mathrm{out}$ denote the input and the output DC powers of the WPT segment, respectively; and where
$\eta_1 \triangleq P_\mathrm{tx}/P_\mathrm{in}$,
$\eta_2 \triangleq P_\mathrm{erp}/P_\mathrm{tx}$,
$\eta_3 \triangleq P_\mathrm{inc}/P_\mathrm{erp}$,
$\eta_4 \triangleq P_\mathrm{rx}/P_\mathrm{inc}$
and
$\eta_5 \triangleq P_\mathrm{out}/P_\mathrm{rx}$,
as shown in Fig.~\ref{Fig_WPT}. Alternatively, $\eta_\mathrm{e2e}$ can be decomposed as the product of the transmit efficiency ($\eta_1 \times \eta_2$), the free-space propagation efficiency ($\eta_3$) and the receive efficiency ($\eta_4 \times \eta_5$).

In the state-of-the-art, in order to improve the end-to-end power transfer efficiency, researchers' efforts were focused on enhancing the transmit efficiency, the receive efficiency, or both. To obtain higher transmit efficiency, the transmit antenna array must be designed such that the side lobes of its beam pattern are reduced to the lowest acceptable level and that its main lobe keeps spillover losses to a minimum. To this end, electronically steered phased arrays with retrodirective beamforming have taken the place of traditional horn antennas and emerged as the most reliable technique to guarantee an accurate beam steering in WPT applications. On the other hand, attaining higher receive efficiency was attempted through the design of high-performance rectifying antennas (i.e. rectennas), which convert the incident RF power back to DC. A comprehensive survey on the design of transmit and receive antenna arrays dedicated to WPT applications can be found in \cite{Massa13PIEEE06}. It is noteworthy that omnidirectional antennas must be avoided in WPT applications in order to improve the power transfer efficiency and, in particular, decrease the risks of human exposure to electromagnetic fields.

Despite the efforts to improve the transmit efficiency ($\eta_1 \times \eta_2$) and/or the receive efficiency ($\eta_4 \times \eta_5$), the main determinant of the end-to-end power transfer efficiency remains $\eta_3$, i.e. the free-space propagation efficiency. In fact, recalling the well-known Friis transmission equation, for an effective radiated power (ERP) $P_\mathrm{erp}$ at a power beacon (defined as the product of transmit power and antenna gain), the incident power $P_\mathrm{inc}$ at a receive antenna array is exponentially decayed with the Euclidean distance between the power beacon and the power receiver (cf. Fig.~\ref{Fig_ReceiverModel}). Without loss of generality, setting the path-loss exponent to the urban environment value of 4, the power transfer efficiency can readily be shown to be much smaller than $1\%$, given that the distance between the beacon and the receiver is on the order of ten meters. Therefore, improving the power transfer efficiency is a prerequisite for the success of far-field WPT.

\subsection{Contributions and Organization of the Work}
In this paper, we exploit multi-user scheduling, which was proven to be effective in improving data transmission efficiency in wireless communication systems, to improve the power transfer efficiency in propagation space (i.e. $\eta_3$ shown in Fig.~\ref{Fig_WPT}) by utilizing the channel variety among the UEs. Moreover, since the rectifier is the core component of a power receiver, its non-linearity effect is  accounted for and the output DC power, $P_\mathrm{out}$, shown in Fig.~\ref{Fig_WPT}, is quantified. For simplicity, the receive antenna array is assumed ideal and, thus, $\eta_4$ is normalized to unity.

To study the effect of channel variety among UEs on the efficiency of far-field power transfer, we first look into the {\it homogeneous} scenario where UEs, described as symmetric in this case, have a similar distance from the power beacon, and we analyze the time-average output DC power when either round-robin scheduling or opportunistic scheduling is implemented for the power transfer to the UEs. Then, we investigate the {\it heterogeneous} scenario with asymmetric UEs, which are uniformly distributed in the considered annulus, and analyze the average output DC power with the said scheduling strategies. The results of this study reveal the effects of the inner and outer boundaries of the system coverage, multi-user scheduling and channel variety among UEs, on the power transfer efficiency of the WPT segment in an out-band system. In particular, three major contributions of this paper are:
\begin{enumerate}
\item Implementing round-robin scheduling, the channel variety among asymmetric UEs is demonstrated to be equivalent to a distance scaling factor, compared to the case with symmetric UEs. Also, the said factor is shown to be determined by the size of the inner and outer boundaries of the power beacon coverage and by the path-loss exponent;
\item Performing opportunistic scheduling while considering $N$ symmetric UEs, the power scaling law is derived and reveals that the scheduling gain is $\ln{N}$, compared to the power transfer efficiency pertaining to round-robin scheduling;
\item Applying the same opportunistic scheduling as above among asymmetric UEs, the power scaling law is analytically attained and discloses that the scheduling gain is as large as $N$ (number of UEs), compared to the power transfer efficiency pertaining to round-robin scheduling. Therefore, the channel variety among UEs can be exploited to improve the far-field power transfer efficiency significantly.
\end{enumerate}

To detail the paper's contributions, the remainder of the manuscript is organized as follows. Section~\ref{Section:SystemModel} describes the system and the channel models. In Section~\ref{Section:EnergyReceiver}, the inner structure of the UE power receiver is sketched and the amount of instantaneous output DC power at the receiver is formulated. Section~\ref{Section:homogeneousCase} is devoted to the scenario with symmetric UEs and analyzes the time-average output DC power and power transfer efficiency with different scheduling policies. Then, Section~\ref{Section:heterogeneousCase} focuses on the scenario with asymmetric UEs. Simulation results and discussions are presented in Section~\ref{Section:Simulation&Discussion}. Concluding remarks are provided in Section~\ref{Section:Conclusion}, followed by mathematical tools and derivations relegated to appendices. In particular, the third contribution mentioned above was reported in the accompanying conference version \cite{XiaGlobecom14}.

{\bf Notation}: For a complex variable $x$, operators $\Re\{x\}$, $\Im\{x\}$, $|x|$ and $\arg\{x\}$ denote its real part, imaginary part, amplitude and phase, respectively. $\mathbb{E}\{x\}$ indicates the statistical expectation of real random variable $x$. Constant $j = \sqrt{-1}$ denotes the imaginary unit. The functions $\Gamma(x) = \int_0^\infty{t^{x-1}e^{-t}\,\mathrm{d}t}$ and $\Gamma(a,\, x) = \int_x^\infty{t^{a-1}e^{-t}\,\mathrm{d}t}$, $\forall a, x > 0$, refer to the Gamma function and the complementary incomplete Gamma function, respectively. The Landau notations $f(x)=\it{O}\left(g(x)\right)$ and $f(x) \sim g(x)$ are defined as $\lim_{x \to \infty}|f(x)/g(x)| < \infty$ and $\lim_{x \to \infty}f(x)/g(x) = 1$, respectively. Finally, $\overline{P_x^y}$ stands for the time-average output DC power with respect to the joint states $x$ and $y$, where $x \in \{\mathrm{rrs}, \rm{os}\}$ refers to the scheduling policy (`$\mathrm{rrs}$' for round-robin and `$\rm{os}$' for opportunistic) and $y \in \{\rm{hom}, \rm{het}\}$ to the status of UEs (`$\rm{hom}$' for the homogeneous case and `$\rm{het}$' for the heterogeneous case).

\section{System and Channel Models}
\label{Section:SystemModel}
\begin{figure}[!t]
\centering
\includegraphics [width=1.8in, clip, keepaspectratio]{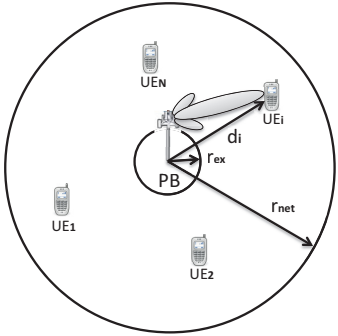}
\caption{The WPT segment in the simultaneous power and data transfer system illustrated in Fig.~\ref{Fig_ReceiverModel}, where the power beacon (PB) in the center of the circular area is steering power to the $i^\mathrm{th}$ user terminal, $\mathrm{UE}_i$, $i \in [1, N]$. All terminals are located in the annulus of inner radius $r_\mathrm{ex}$ and outer radius $r_\mathrm{net}$.}
\label{Fig_Scheduling}
\end{figure}

As depicted in Fig.~\ref{Fig_Scheduling}, we consider the WPT segment in an out-band system where the power beacon steers power to the $i^\mathrm{th}$ user equipment, $\mathrm{UE}_i$, $\forall i \in [1, N]$, through a directional antenna array. The power beacon is located in the center of a circular coverage with radius $r_\mathrm{net}$, with an exclusion zone of radius $r_\mathrm{ex}$ around it such that all UEs are in the far-field of the power beacon. The exclusion zone can be seen as a prohibited zone where users are not allowed for human safety reasons due to the high radiated power of power beacons (more details on the safety issue will be examined in Remark 4).

At the power beacon, the radiated RF signal at time slot $t$ is expressed as
\begin{equation} \label{TransmittedSymbol}
s(t) = \sqrt{2P_\mathrm{erp}}\,\Re\left\{x(t) \, e^{j2{\pi}ft}\right\},
\end{equation}
where $P_\mathrm{erp}=\mathbb{E}\{s^2(t)\}$ denotes the average effective radiated power at the transmit antenna array, $f$ refers to the carrier frequency, and $x(t)$ is a complex baseband signal of bandwidth $B$ Hz and unit power (i.e. $|x(t)|^2 = 1$). Also, it is assumed that wireless channels experienced by $s(t)$ keep flat during each transmission slot.

A basic wireless channel model between the power beacon and receiver $\mathrm{UE}_i$ consists of  large-scale path loss and small-scale multi-path fading. Mathematically, the instantaneous multiplicative channel gain $G_i(t)$ between the power beacon and $\mathrm{UE}_i$ at time slot $t$ is given by
\begin{equation} \label{ChannelModel}
G_i(t) = \beta{d_i^{-\alpha}(t)}|h_i(t)|^2,
\end{equation}
where $\beta > 0$ is a constant scaling factor,\footnote{In general, parameter $\beta$ in Eq.~\eqref{ChannelModel} stands for the log-normal shadowing effect introduced by objects obstructing the propagation path between the power beacon and the UE. This effect influences the local-mean powers at the UE, that is, short-term averages to remove fluctuations due to multi-path fading \cite[Section 2.4]{Stuber02}, which is accounted for by the parameter $\sigma_h^2$ defined immediately after Eq.~\eqref{CDF-Rayleigh} of the paper.} $d_i(t) \in [r_\mathrm{ex},~r_\mathrm{net}]$ is the distance between the power beacon and $\mathrm{UE}_i$, $\alpha \ge 2$ denotes the path-loss exponent, and $h_i(t)$ stands for the  complex channel coefficient.

\section{Power Receiver}
\label{Section:EnergyReceiver}
\begin{figure*}[!t]
\centering
\includegraphics [width=5in, clip, keepaspectratio]{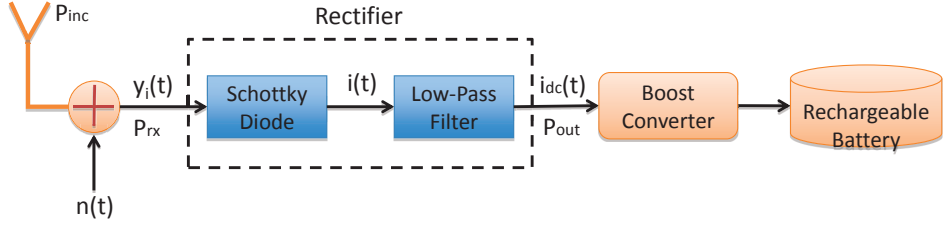}
\caption{Block diagram of a typical rectifier-based power receiver.}
\vspace{-10pt}
\label{Fig_EnergyReceiver}
\end{figure*}

With respect to the receiver, Fig. \ref{Fig_EnergyReceiver} illustrates a typical rectifier-based power receiver, where the incident RF power, $P_\mathrm{inc}$, is first converted by the rectifier into direct current (DC). Then, a boost converter is cascaded to provide the required step-up from typical rectenna voltage (usually, tens to hundreds millivolts) to typical battery voltage (usually, 2 to 4 volts). As shown in the dashed block of Fig.~\ref{Fig_EnergyReceiver}, the rectifier is composed of a Schottky diode cascaded with a low-pass filter (LPF), thanks to the low power-loss and the fast switching/recovery time of Schottky diode, compared with ordinary P-N junction diodes.

In light of \eqref{TransmittedSymbol} and \eqref{ChannelModel}, the received signal at the power receiver of $\mathrm{UE}_i$ can be shown to be expressed as
\begin{equation} \label{ReceivedSignal-1}
y_i(t) = \sqrt{2P_\mathrm{erp}G_i(t)}\,\Re\left\{x(t) \, e^{j(2{\pi}ft+\theta_i(t))}\right\} + n_i(t),
\end{equation}
where $\theta_i(t) \in [0, 2\pi)$ and $n_i(t)$ denote the phase shift introduced by the propagation channel and the additive white Gaussian noise (AWGN) at the receive antenna of $\mathrm{UE}_i$, respectively. Without loss of generality, $n_i(t)$ is assumed to have zero mean and variance $\delta^2$. Notice that, for the considered far-field WPT segment, the average effective radiated power, $P_\mathrm{erp}$, is generally much larger than thermal noise variance, $\delta^2$. As a result, \eqref{ReceivedSignal-1} can be well approximated by using
\begin{equation} \label{ReceivedSignal-1-a}
y_i(t) \approx \sqrt{2P_\mathrm{erp}G_i(t)}\,\Re\left\{x(t) \, e^{j\left(2{\pi}ft+\theta_i(t)\right)}\right\}.
\end{equation}
Then, for an easier mathematically tractable form of $y_i(t) $, \eqref{ReceivedSignal-1-a} can be reformulated as:
\begin{eqnarray}
y_i(t)
&  \approx   & \sqrt{2}\,\Re\left\{\sqrt{P_\mathrm{erp}G_i(t)} \, x(t)e^{j(2{\pi}ft+\theta_i(t))} \right\}				 \nonumber\\
&  =   & \sqrt{2}\,\Re\left\{\sqrt{P_\mathrm{erp}G_i(t)} \, e^{j[\theta_i(t)+\arg(x(t))]}e^{j2{\pi}ft} \right\}			 \nonumber\\
&  =   & \sqrt{2}\,A(t)\cos\left[2{\pi}ft+\phi(t)\right],                                        										 \label{ReceivedSignal-2}
\end{eqnarray}
where the amplitude $A(t)$ and the phase $\phi(t)$ are given by
\begin{equation}
A(t) = \sqrt{y_I^2(t) + y_Q^2(t)},
\end{equation}
\begin{equation}
\phi(t) = \arctan{\frac{y_Q(t)}{y_I(t)}},
\end{equation}
with
\begin{equation} \label{Inphase}
y_I(t) \triangleq \sqrt{P_\mathrm{erp}G_i(t)}\,\cos{[\theta_i(t)+\arg(x(t))]},
\end{equation}
\begin{equation} \label{Quadratic}
y_Q(t) \triangleq \sqrt{P_\mathrm{erp}G_i(t)}\,\sin{[\theta_i(t)+\arg(x(t))]}.
\end{equation}

Since the current-voltage characteristic of a Schottky diode can be described by the well-known diode law, in view of the sinusoidal input voltage given by \eqref{ReceivedSignal-2}, the output current of the Schottky diode in Fig.~\ref{Fig_EnergyReceiver} can be readily given by
\begin{equation} \label{CVcharacteristic-1}
i(t) = I_s\left(\exp\left({\displaystyle \frac{y_i(t)}{\rho V_T}}\right)-1\right),
\end{equation}
where $I_s$ denotes the reverse saturation current of the diode, $\rho$ is the quality factor of the diode and $V_T$ refers to the thermal voltage. The value of $\rho$ typically varies from 1 to 2 depending upon the fabrication process and semiconductor material and, usually, is assumed to be approximately equal to~1. On the other hand, $V_T = kT/e$, where $k$ is the Boltzmann constant, $T$ is the working temperature in Kelvin, and $e$ is the magnitude of the electronic charge. Subsequently, by using the Taylor series expansion of the exponential function in \eqref{CVcharacteristic-1}, the output current can be rewritten as
\begin{equation} \label{CVcharacteristic-2}
 i(t) = \sum_{k=1}^\infty{\frac{I_s}{(\rho V_T)^k}\,y_i^k(t)}.
\end{equation}

After passing through the LPF immediately after the Schottky diode, the harmonic components at $kf$ ($k \ge 1$) of $i(t)$ shown in \eqref{CVcharacteristic-2} are removed and only the DC component  remains at the output of the rectifier (cf. Fig.~\ref{Fig_EnergyReceiver}). Therefore, by ignoring the high-order ($k > 2$) terms in \eqref{CVcharacteristic-2} and performing some  trigonometric manipulations, the DC output of the rectifier can be readily given by
 \begin{equation}
i_\mathrm{dc}(t)
= \frac{I_s}{(\rho V_T)^2}\,A^2(t)	
= \frac{I_s P_\mathrm{erp}}{(\rho V_T)^2}\,G_i(t).		\label{DC-2}
 \end{equation}

With the input current $i_\mathrm{dc}(t)$, the converted power to be stored in the rechargeable battery is in general linearly proportional to the value of $i_\mathrm{dc}(t)$, with a conversion coefficient $0 < \xi \le 1$ ($\xi > 0.85$ in practice \cite{Brown84TMTT09}). Thus, the instantaneous output DC power $P_{\mathrm{out}-i}(t)$ at $\mathrm{UE}_i$ during transmission slot $t$, in the unit of watt or equivalently joule/sec, is given by
\begin{equation} \label{HarvestedEnergy}
P_{\mathrm{out}-i}(t) = \xi \, i_\mathrm{dc}(t) = \frac{{\xi}I_{s}P_\mathrm{erp}}{(\rho V_T)^2}\,G_i(t) =c\beta{d_i^{-\alpha}(t)}|h_i(t)|^2,
\end{equation}
where $c \triangleq {\xi}I_{s}P_\mathrm{erp}/(\rho V_T)^2$. Clearly, \eqref{HarvestedEnergy} demonstrates that the instantaneous output DC power at $\mathrm{UE}_i$ during each transmission slot is proportional to the path loss $d_i^{-\alpha}(t)$, which is the fundamental reason why the WPT efficiency degrades significantly with the propagation distance.

Recalling the five subsystems of Fig.~\ref{Fig_WPT}, it is noteworthy that the mathematical model starting from \eqref{TransmittedSymbol} to \eqref{HarvestedEnergy} captures the physical processes where the signals propagate over space and are processed by the rectifier in a power receiver illustrated in Fig.~\ref{Fig_EnergyReceiver}. Hence, the power transfer efficiency described hereafter is equivalent to the product of $\eta_3$ and $\eta_5$ whereas $\eta_1$, $\eta_2$ and $\eta_4$ (cf. Fig.~\ref{Fig_WPT}) are not accounted for.\footnote{For more details on the way to improve the subsystems' efficiencies $\eta_1$, $\eta_2$ and $\eta_4$, the interested reader is referred to \cite{Massa13PIEEE06, OliveriIWPT12} and references therein.}

In order to investigate the effect of the distance $d_i$ on the power transfer efficiency in a multi-user context, in the sequel we consider two different scenarios in conjunction with different multi-user scheduling strategies for the power transfer from the beacon.

\begin{remark}[Rectifier and rectifying theory]
The typical rectifier-based power receiver shown in Fig.~\ref{Fig_EnergyReceiver} is a normal wave detector suitable for relatively weak power receiver and/or energy harvester. It enables mathematical tractability without any side effect on the major findings of the paper. For medium and high power-consumption applications, a single-shunt full-wave rectifier can be used, which consists of a diode and a capacitor connected in parallel to a ${\lambda_{g}}/4$ distributed line, with $\lambda_{g}$ being the effective wavelength of an input radiowave \cite[Fig. 3.67]{Shinohara14}. For the general rectifying theory and the state-of-the-art of various rectennas, the interested reader is referred to \cite{Shinohara14}.
\end{remark}

\section{Scenario A: Homogeneous Users}
\label{Section:homogeneousCase}
In this scenario, UEs are assumed to be uniformly distributed along a circle centered by the power beacon or located closely to each other, widely known as clustered users (e.g. several tablets on a table while the PB is adsorbed onto the ceiling above, or users in a small coffee shop). In such a case, UEs have almost the same distance from the power beacon and, thus, they get the same average power as per \eqref{HarvestedEnergy}.  Accordingly, for ease of presentation, in this section the distance $d_i(t)$ is abbreviated as $d$, regardless of the user index ($i$) or the time variable ($t$). Actually, if we set $d=1$, this idealized scenario reduces to the case where the path-loss effect is ignored, as usually assumed impractically in the open literature related to WPT.

Next, in order to study the effect of path-loss on the WPT efficiency and, in particular, provide a performance benchmark for later development, two different scheduling strategies among symmetric UEs are considered and their corresponding average output DC powers are analyzed.

\subsection{Round-Robin Scheduling}
\label{ScenarioA-RoundRobin}
In round-robin scheduling, UEs take periodic turns to access the medium of service. Under this policy, all UEs get an equal share of the available resource, such as time, and thus have the same performance.

By recalling that $d_{i}(t) \equiv d$ for all UEs in the homogeneous case, the instantaneous output DC power at $\mathrm{UE}_i$, given by \eqref{HarvestedEnergy}, can be rewritten as
\begin{equation}
\label{HarvestEnergy-2}
P_{\mathrm{out}-i}(t) = c\beta{d^{-\alpha}}|h_{i}(t)|^2.
\end{equation}
Assuming Rayleigh model for the effect of multi-path fading, the term $|h_{i}(t)|^2$ in \eqref{HarvestEnergy-2} has an exponential distribution such that the probability density function (PDF) and the cumulative density function (CDF) can be expressed as follows:
\begin{equation} \label{PDF-Nakagami}
f_{|h_{i}(t)|^2}(x)
= \frac{1}{\sigma_h^2}\exp\left(-\frac{x}{\sigma_h^2}\right),
\end{equation}
\begin{equation}
\label{CDF-Rayleigh}
F_{|h_{i}(t)|^2}(x)
= 1-\exp\left(-\frac{x}{\sigma_h^2}\right),
\end{equation}
where $\sigma_h^2 \triangleq \mathbb{E}\{|h_{i}(t)|^2\}, \forall i \in [1, N]$, denotes the average small-scale multi-path gain. By virtue of \eqref{HarvestEnergy-2}--\eqref{CDF-Rayleigh}, it is easy to derive the PDF and the CDF of $P_{\mathrm{out}-i}(t)$, namely,
\begin{equation}
\label{PDF-Energy-Case1}
f_{P_{\mathrm{out}-i}(t)}(x)
= \frac{d^\alpha}{c\beta\sigma_h^2}\exp\left(-\frac{d^\alpha}{c\beta\sigma_h^2} \, x\right),
\end{equation}
\begin{equation}
\label{CDF-Energy-Case1}
F_{P_{\mathrm{out}-i}(t)}(x)
= 1-\exp\left(-\frac{d^\alpha}{c\beta\sigma_h^2} \, x\right).
\end{equation}

When round-robin scheduling is implemented, each UE gets the same average output DC power. In view of \eqref{PDF-Energy-Case1}, this power value can be easily computed as
\begin{equation}
\label{AverageEnergy-Case1}
\overline{P_{\mathrm{rrs}}^{\mathrm{hom}}}
= \int_0^\infty{x f_{P_{\mathrm{out}-i}(t)}(x) }\mathrm{d}x
= c\beta\sigma_h^2 \, {d^{-\alpha}},
\end{equation}
which implies that the average output DC power increases linearly with larger multi-path gain ($\sigma_h^2$) but decreases exponentially with the transmission distance ($d$), as expected. Although the result in \eqref{AverageEnergy-Case1} is straightforward, it will serve as a benchmark for the subsequent complex cases.

\subsection{Opportunistic Scheduling}
\label{ScenarioA-Opportunistic}
Although round-robin scheduling policy ensures absolute fairness among UEs, in terms of access to the system resources, it is not very efficient. For instance, for UEs which experience deep fading, the amount of instantaneous output DC power can be extremely small. In order to enhance the power transfer efficiency, opportunistic scheduling, which was originally proposed to improve spectral efficiency, can be exploited \cite{KnoppICC95}. In this case, assuming that a UE cannot get fully charged when scheduled, the UE with the maximum instantaneous output DC power is chosen to be charged during each transmission slot. To this end, a transmission slot is divided into two sub-slots: one training sub-slot and one power-transfer sub-slot. During the training sub-slot, each UE sends its received power level back to the power beacon by a reliable feedback mechanism, and the power beacon chooses the one which has the highest power level to be charged in the following power-transfer sub-slot. With such a multi-user scheduling strategy, the index of the chosen UE is given by

\begin{equation}	\label{Eq:ChosenUser}
\hat{\imath} = \arg\max\limits_{i = 1, \, \cdots, \, N}{P_{\mathrm{out}-i}(t)},
\end{equation}
where $P_{\mathrm{out}-i}(t)$ denotes the instantaneous received power of $\mathrm{UE}_i$, defined in \eqref{HarvestEnergy-2}. By recalling the results of order statistics theory, the CDF and the PDF of the maximum output DC power $P_{\mathrm{out}-\hat{\imath}}(t)$ of the chosen $\mathrm{UE}_{\hat{\imath}}$ is readily shown as
\begin{equation}	\label{Eq:CDFMaximuEnergy}
F_{P_{\mathrm{out}-\hat{\imath}}(t)}(x) = F_{P_{\mathrm{out}-i}(t)}^N(x),
\end{equation}
\begin{equation}	\label{Eq:PDFMaximuEnergy}
f_{P_{\mathrm{out}-\hat{\imath}}(t)}(x) = N f_{P_{\mathrm{out}-i}(t)}(x) F_{P_{\mathrm{out}-i}(t)}^{N-1}(x),
\end{equation}
where $f_{P_{\mathrm{out}-i}(t)}(x)$ and $F_{P_{\mathrm{out}-i}(t)}(x)$ are given by \eqref{PDF-Energy-Case1} and \eqref{CDF-Energy-Case1}, respectively. Accordingly, the average output DC power can be computed by
\begin{equation}	\label{Eq:AverageEnergy-Case-3}
\overline{P_{\mathrm{os}}^{\mathrm{hom}}}
 = N\int_0^\infty{x f_{P_{\mathrm{out}-i}(t)}(x) F_{P_{\mathrm{out}-i}(t)}^{N-1}(x)}\,\mathrm{d}x.
\end{equation}

Although substituting \eqref{PDF-Energy-Case1} and \eqref{CDF-Energy-Case1} into \eqref{Eq:AverageEnergy-Case-3} and performing some mathematical manipulations leads to a closed-form expression, the resultant formula is very complex and has to be evaluated numerically. In order to offer illuminating insights into the output DC power, we instead derive the limiting distribution of the maximum output DC power $P_{\mathrm{out}-\hat{\imath}}(t)$, as $N \to \infty$. It is noteworthy that the limiting distribution cannot be obtained by directly applying $N \to \infty$ in \eqref{Eq:CDFMaximuEnergy} since, for any $F_{P_{\mathrm{out}-\hat{\imath}}(t)}(x) < 1$, \eqref{Eq:CDFMaximuEnergy} reduces to $0$ as $N \to \infty$ and, thus, the CDF given by \eqref{Eq:CDFMaximuEnergy}  is a degenerate distribution.

In the following lemma, the asymptotic theory of extreme order statistics is exploited to attain a non-degenerate distribution for $P_{\mathrm{out}-\hat{\imath}}(t)$ and, then, the power scaling law is derived.

\begin{lemma}[The limiting distribution of the maximum instantaneous output DC power among symmetric UEs]
\label{Lemma1}
If the PDF and CDF of the output DC power at $\mathrm{UE}_i$, $\forall i \in [1, N]$, are respectively given by \eqref{PDF-Energy-Case1}  and \eqref{CDF-Energy-Case1}, then as the total number of UEs, $N$, becomes large asymptotically, the limiting distribution of the maximum instantaneous output DC power $P_{\mathrm{out}-\hat{\imath}}(t)$ is of Gumbel type, namely,
\begin{equation} \label{Eq.LimitingCase3}
\lim\limits_{N \to \infty}F_{P_{\mathrm{out}-\hat{\imath}}(t)}(a_1+b_{1}\,x) = \exp\left(-e^{-x}\right),
\end{equation}
where the positioning parameter $a_1$ and the scaling factor $b_1$ are given by
\begin{equation} \label{Eq.LimitingCase3-a}
a_1 = c \beta\sigma_h^2\, d^{-\alpha} \ln{N},
\end{equation}
\begin{equation}  \label{Eq.LimitingCase3-b}
b_1 = c \beta\sigma_h^2\, d^{-\alpha}.
\end{equation}
\end{lemma}
\begin{IEEEproof}
The proof is provided in Appendix~\ref{AppendixA}.
\end{IEEEproof}

Recalling the fact that the mean of the Gumbel function is Euler's constant, i.e. $0.5772 \cdots$ \cite[p.~298]{David03}, with the resulting Lemma~\ref{Lemma1}, it is straightforward that the average output DC power can be computed as
\begin{equation}
\overline{P_{\mathrm{os}}^{\mathrm{hom}}}
=  c \beta\sigma_h^2\, d^{-\alpha}(\ln{N} + 0.5772\cdots) \label{AverageEnergy-Case3}.
\end{equation}
In light of \eqref{AverageEnergy-Case3}, we immediately have the following theorem.

\begin{theorem}[Power scaling law for symmetric UEs]
\label{Thoerem-homogeneousCase}
When there are $N$ UEs which have the same distance ($d$) from the power beacon, if the UE with the maximum instantaneous output DC power is chosen to be charged during each transmission slot, then as $N$ becomes large asymptotically, the average output DC power at each UE scales as
\begin{equation}
\label{ScalingLaw-homogeneousCase}
\overline{P_{\mathrm{os}}^{\mathrm{hom}}}
\sim c \beta \sigma_h^2\, d^{-\alpha}\ln{N}.
\end{equation}
\end{theorem}

\begin{IEEEproof}
When the value of $N$ becomes large asymptotically, ignoring the Euler's constant in the parentheses of \eqref{AverageEnergy-Case3} yields the desired \eqref{ScalingLaw-homogeneousCase}.
\end{IEEEproof}

By definition, the power transfer efficiency pertaining to opportunistic scheduling is given by $\overline{P_{\mathrm{os}}^{\mathrm{hom}}}/P_{\mathrm{erp}}$ while that pertaining to round-robin scheduling is expressed as $\overline{P_{\mathrm{rrs}}^{\mathrm{hom}}}/P_{\mathrm{erp}}$. Accordingly, the ratio of their power transfer efficiencies can be shown as $\overline{P_{\mathrm{os}}^{\mathrm{hom}}}/\overline{P_{\mathrm{rrs}}^{\mathrm{hom}}}$. Consequently, by comparing \eqref{ScalingLaw-homogeneousCase} with \eqref{AverageEnergy-Case1}, we conclude the following corollary.
\begin{corollary}[Power transfer efficiency for symmetric UEs]
\label{Corolloary-fixedCase}
The opportunistic scheduling among $N$ symmetric UEs yields a scheduling gain of~$\ln{N}$ in terms of the power transfer efficiency, compared with the round-robin scheduling.
\end{corollary}

\begin{remark}[On the multi-user scheduling in the context of power transfer]
In the context of power transfer, the received-power based multi-user scheduling, given by Eq.~\eqref{Eq:ChosenUser}, is essentially equivalent to the channel state information (CSI) based multi-user scheduling in the context of data exchange in conventional communication systems. However, unlike the data receiver where the CSI can be obtained by the built-in channel estimation unit, it is not necessary for the power receiver to estimate CSI and, thus, the feedback information from the power receiver to its target power beacon is only the received power level. On the other hand, due to the relatively very short duration of the training sub-slot in comparison with the subsequent power-transfer sub-slot, the energy obtained at each terminal during the training sub-slot is negligible.
\end{remark}

\section{Scenario B: Heterogeneous Users}
\label{Section:heterogeneousCase}
When UEs are uniformly distributed as depicted in Fig.~\ref{Fig_Scheduling}, the distance $d_{i}(t)$ between $\mathrm{UE}_i$ and the power beacon is a random variable with respect to the user index $i$ and time variable $t$. In such a case, the average output DC power of different UEs vary dramatically with their respective distances from the power beacon; a scenario referred to as asymmetric UEs. In this section, taking the distance variety into account, we investigate the power transfer efficiency of the system by means of the time-average output DC power. First, we start with the scheduling according to the round-robin policy.

\subsection{Round-Robin Scheduling}
\label{ScenarioB-RoundRobin}
When UEs are uniformly distributed in the annulus of inner radius $r_\mathrm{ex}$ and outer radius $r_\mathrm{net}$ (cf. Fig.~\ref{Fig_Scheduling}), the distance $d_{i}(t)$ of $\mathrm{UE}_i$ from the power beacon is a random variable and its CDF and PDF are respectively given by
\begin{equation} \label{CDF-Distance}
F_{d_{i}(t)}(x)
= \begin{cases}
\displaystyle \frac{x^2 - r_\mathrm{ex}^2}{r_\mathrm{net}^2 - r_\mathrm{ex}^2},	& \text{if } r_\mathrm{ex}\le x \le r_\mathrm{net}, \\
0,	& \text{otherwise};
\end{cases}
\end{equation}
\begin{equation} \label{PDF-Distance}
f_{d_{i}(t)}(x)
= \begin{cases}
\displaystyle \frac{2x}{r_\mathrm{net}^2 - r_\mathrm{ex}^2},	& \text{if } r_\mathrm{ex}\le x \le r_\mathrm{net}, \\
0,	& \text{otherwise}.
\end{cases}
\end{equation}

In light of \eqref{PDF-Nakagami}, \eqref{CDF-Rayleigh},  \eqref{CDF-Distance} and \eqref{PDF-Distance}, and performing some mathematical manipulations (details provided in Appendix~\ref{Appendix-B}), the CDF of the instantaneous output DC power given by \eqref{HarvestedEnergy} can be found to be expressed as
\begin{eqnarray}	\label{CDF-Energy-Case2}
\lefteqn{F_{P_{\mathrm{out}-i}(t)}(x)}   \nonumber \\
&  =  & 1- \frac{1}{r_\mathrm{net}^2 - r_\mathrm{ex}^2} \left[r_\mathrm{net}^2 \exp\left(-r_2\,x\right) - r_\mathrm{ex}^2 \exp\left(-r_1\,x\right)\right]		\nonumber \\
&     &{} - \frac{1}{r_\mathrm{net}^2 - r_\mathrm{ex}^2} \left(\frac{c\beta\sigma_h^2}{x} \right)^{\frac{2}{\alpha}}		  \nonumber \\
&     &{}\times \left[\Gamma\left(1+\frac{2}{\alpha}, \, r_1\,x\right) - \Gamma\left(1+\frac{2}{\alpha}, \, r_2\,x\right)\right],
\end{eqnarray}
where $r_1 \triangleq r_\mathrm{ex}^\alpha/(c \beta\sigma_h^2)$ and $r_2 \triangleq r_\mathrm{net}^\alpha/(c \beta\sigma_h^2)$.

When round-robin scheduling is adopted, all UEs get the same average output DC power and, with the resultant \eqref{CDF-Energy-Case2}, it can be computed as
\begin{small}
\begin{eqnarray}
\overline{P_{\mathrm{rrs}}^{\mathrm{het}}}
&  =  & \int\limits_0^\infty{\left[1-F_{P_{\mathrm{out}-i}(t)}(x)\right]}\mathrm{d}x		\nonumber\\
&  =  & \frac{1}{r_\mathrm{net}^2 - r_\mathrm{ex}^2} \int\limits_0^\infty{[r_\mathrm{net}^2 \exp\left(-r_2\,x\right) - r_\mathrm{ex}^2 \exp\left(-r_1\,x\right)]}\,\mathrm{d}x		\nonumber\\
&      &{}+ \frac{\left(c\beta\sigma_h^2\right)^{\frac{2}{\alpha}}}{r_\mathrm{net}^2 - r_\mathrm{ex}^2} 		 \nonumber \\
&      &{}\times \int\limits_0^\infty{\hspace{-2pt} x^{-\frac{2}{\alpha}} \left[\Gamma\left(1+\frac{2}{\alpha}, \, r_1\,x\right) - \Gamma\left(1+\frac{2}{\alpha}, \, r_2\,x\right)\right]} \,\mathrm{d}x		\nonumber\\
&  =  & \frac{1}{r_\mathrm{net}^2 - r_\mathrm{ex}^2}\left(\frac{r_\mathrm{net}^2}{r_2}-\frac{r_\mathrm{ex}^2}{r_1}\right)    \nonumber \\
&      &{}+\frac{\alpha \, (c \beta\sigma_h^2)^{\frac{2}{\alpha}}}{(\alpha-2)(r_\mathrm{net}^2 - r_\mathrm{ex}^2)}\left(r_1^{\frac{2}{\alpha}-1}-r_2^{\frac{2}{\alpha}-1}\right)		 \label{AverageEnergy-Case2-a} \\
&  =  & \frac{2c\beta\sigma_h^2}{(\alpha-2)(r_\mathrm{net}^2 - r_\mathrm{ex}^2)}\left(r_\mathrm{ex}^{2-\alpha} - r_\mathrm{net}^{2-\alpha}\right),		 \label{AverageEnergy-Case2}
\end{eqnarray}
\end{small}
\hspace{-6pt} where \cite[Eq.~(6.455.1)]{Gradshteyn07} was exploited to derive \eqref{AverageEnergy-Case2-a}. Thus, \eqref{AverageEnergy-Case2} establishes the relationship between the average output DC power and the system parameters in a very simple and explicit way. Furthermore, it is general in practice that $r_\mathrm{net} \gg r_\mathrm{ex}$ and, thus, \eqref{AverageEnergy-Case2} can be approximated as follows:
\begin{eqnarray}
\overline{P_{\mathrm{rrs}}^{\mathrm{het}}}
&  \approx  & \frac{2 c \beta\sigma_h^2}{\alpha-2}\left(\frac{r_\mathrm{ex}}{r_\mathrm{net}}\right)^2 r_\mathrm{ex}^{-\alpha}		\label{AverageEnergy-Case2-Approx-1} \\
&      =         & \underbrace{c\beta\sigma_h^2 \, { r_\mathrm{ex}^{-\alpha}}}_{T_1} \,
\underbrace{ \frac{2}{\alpha-2}\left(\frac{r_\mathrm{ex}}{r_\mathrm{net}}\right)^2}_{T_2}.	 \label{AverageEnergy-Case2-Approx-2}
\end{eqnarray}

By comparing \eqref{AverageEnergy-Case2-Approx-2} with \eqref{AverageEnergy-Case1}, it is observed that: 1)  The term $T_1$ of \eqref{AverageEnergy-Case2-Approx-2} corresponds to the maximum average output DC power of the symmetric-user case with the shortest distance $r_\mathrm{ex}$ from the power beacon; 2) The term $T_2$ of \eqref{AverageEnergy-Case2-Approx-2} is a power scaling factor (with respect to $T_1$) that reveals the effect of path-loss variety among UEs on the average DC power.

On the other hand, if we set \eqref{AverageEnergy-Case2-Approx-2} equal to  \eqref{AverageEnergy-Case1}, we may find a symmetric-user case, whereby the users' distance from the power beacon is $\bar{d}$, and users get the same average output DC power (accordingly, the same power transfer efficiency) as in the heterogeneous case. Specifically, $\bar{d}$ is determined by
\begin{equation}
c\beta\sigma_h^2 \, {\bar{d}}^{-\alpha} = c\beta\sigma_h^2 \, {r_\mathrm{ex}^{-\alpha}} \, \frac{2}{\alpha-2}\left(\frac{r_\mathrm{ex}}{r_\mathrm{net}}\right)^2,
\end{equation}
which yields
\begin{equation}
\bar{d} = r_\mathrm{ex} \, \underbrace{\left[\frac{\alpha-2}{2}\left(\frac{r_\mathrm{net}}{r_\mathrm{ex}}\right)^2\right]^{\frac{1}{\alpha}}}_{T_3}.		 \label{DistanceScalingFator}
\end{equation}
Clearly, the term $T_3$ in \eqref{DistanceScalingFator} is a distance scaling factor (with respect to the shortest distance $r_\mathrm{ex}$) which reflects the effect of path-loss variety among UEs on the average DC power. In other words, if round-robin scheduling is adopted, the average output DC power of asymmetric UEs is equivalent to that of the symmetric ones at a distance $r_\mathrm{ex}T_3$ from the power beacon.

\subsection{Opportunistic Scheduling}
\label{ScenarioB-Opportunistic}
When opportunistic scheduling is applied among the asymmetric UEs, it yields higher scheduling gain than the above homogeneous case. To show this, by using again the asymptotic theory of extreme order statistics (but completely different from the proof of Lemma~\ref{Lemma1}), we obtain the following result.
\begin{lemma}[The limiting distribution of the maximum instantaneous output DC power among asymmetric UEs]
\label{Lemma2}
If the CDF of the instantaneous output DC power at $\mathrm{UE}_i$, $\forall i \in [1, N]$, of the heterogeneous scenario is given by \eqref{CDF-Energy-Case2}, then as $N$ becomes large asymptotically, the limiting distribution of the maximum instantaneous output DC power $P_{\mathrm{out}-\hat{\imath}}(t)$ is of Fr\'{e}chet distribution, namely,
\begin{equation} \label{Eq.LimitingCase4}
\lim\limits_{N \to \infty}F_{P_{\mathrm{out}-\hat{\imath}}(t)}(b_{2}\,x) = \exp\left(-x^{-\frac{2}{\alpha}}\right),
\end{equation}
where the scaling factor $b_2$ is given by
\begin{equation}  \label{Eq.LimitingCase4-b}
b_2 = \frac{2cN\beta\sigma_h^2}{\alpha-2}\left(\frac{r_{\rm{ex}}}{r_{\rm{net}}}\right)^2\,r_\mathrm{ex}^{-\alpha}.
\end{equation}
\end{lemma}
\begin{IEEEproof}
The proof is provided in Appendix~\ref{AppendixC}.
\end{IEEEproof}

Since the path-loss exponent $\alpha \ge 2$ holds in practice, the exponent of $x$ in the right-hand side of \eqref{Eq.LimitingCase4} is no less than $-1$, i.e. $-\frac{2}{\alpha} \ge -1$. In this case, the mean of the Fr\'{e}chet distribution approaches infinity, i.e. the integral $\int_0^\infty \left[1-\exp\left(-x^{-\frac{2}{\alpha}}\right)\right]\,\mathrm{d}x$ does not converge. Therefore, we cannot use a similar methodology as in Section~\ref{ScenarioA-Opportunistic} to attain the power scaling law. To proceed, we derive the upper and lower bounds on the natural logarithm of the maximum instantaneous output DC power, $P_{\mathrm{out}-\hat{\imath}}(t)$, and then obtain the power scaling law.

Specifically, by virtue of Lemma~\ref{Lemma2}, the CDF of $P_{\mathrm{out}-\hat{\imath}}(t)$ can be explicitly shown as
\begin{small}
\begin{equation} \label{Eq.Theorem2-Derivation-1}
\mathrm{Pr}\left\{P_{\mathrm{out}-\hat{\imath}}(t) \le \frac{2cN\beta\sigma_h^2}{\alpha-2}\left(\frac{r_{\rm{ex}}}{r_{\rm{net}}}\right)^2\,r_\mathrm{ex}^{-\alpha} \, x\right\} = \exp\left(-x^{-\frac{2}{\alpha}}\right),
\end{equation}
\end{small}
\hspace*{-0.135in} for $x > 0$ and $N \to \infty$. By recalling the fact that the logarithm function $\ln{x}$ is a monotonically increasing function with respect to $x > 0$, \eqref{Eq.Theorem2-Derivation-1} can be reformulated as
\begin{small}
\begin{multline} \label{Eq.Theorem2-Derivation-2}
\mathrm{Pr}\left\{\ln{P_{\mathrm{out}-\hat{\imath}}(t)} \le \ln\left(\frac{2c\beta\sigma_h^2}{\alpha-2}\left(\frac{r_{\rm{ex}}}{r_{\rm{net}}}\right)^2\,r_\mathrm{ex}^{-\alpha}\right)+{} \ln{N} + \ln{x}\Biggr\}\right.  \\
= \exp\left(-x^{-\frac{2}{\alpha}}\right),
\end{multline}
\end{small}
\hspace*{-0.135in} for $x > 0$ and $N \to \infty$.

Now, taking $x = \ln{N}$, it is clear that $\lim_{N \to \infty}{\exp\left(-(\ln{N})^{-\frac{2}{\alpha}}\right)}  = 1$ and, thus, \eqref{Eq.Theorem2-Derivation-2} implies
\begin{equation} \label{Eq.Theorem2-Derivation-3}
\ln{P_{\mathrm{out}-\hat{\imath}}(t)}
\le \ln\left(\frac{2c\beta\sigma_h^2}{\alpha-2}\left(\frac{r_{\rm{ex}}}{r_{\rm{net}}}\right)^2\,r_\mathrm{ex}^{-\alpha}\right) + \ln{N} + \ln\ln{N}
\end{equation}
almost surely when $N \to \infty$.

On the other hand, \eqref{Eq.Theorem2-Derivation-2} can be equivalently rewritten as
\begin{small}
\begin{multline} \label{Eq.Theorem2-Derivation-4}
\mathrm{Pr}\left\{\ln{P_{\mathrm{out}-\hat{\imath}}(t)} \ge \ln\left(\frac{2c\beta\sigma_h^2}{\alpha-2}\left(\frac{r_{\rm{ex}}}{r_{\rm{net}}}\right)^2\,r_\mathrm{ex}^{-\alpha}\right) +{} \ln{N} + \ln{x} \Biggr\} \right. \\
= 1-\exp\left(-x^{-\frac{2}{\alpha}}\right),
\end{multline}
\end{small}
\hspace*{-0.110255in} for $x > 0$ and $N \to \infty$. Now, taking $x = 1/\ln{N}$, it is clear that $\lim_{N \to \infty}{\exp\left(-(1/\ln{N})^{-\frac{2}{\alpha}}\right)}  = 0$ and, hence, \eqref{Eq.Theorem2-Derivation-4} implies
\begin{equation} \label{Eq.Theorem2-Derivation-5}
\ln{P_{\mathrm{out}-\hat{\imath}}(t)}
\ge \ln\left(\frac{2c\beta\sigma_h^2}{\alpha-2}\left(\frac{r_{\rm{ex}}}{r_{\rm{net}}}\right)^2\,r_\mathrm{ex}^{-\alpha}\right) + \ln{N} - \ln\ln{N}
\end{equation}
almost surely when $N \to \infty$.

Finally, combining \eqref{Eq.Theorem2-Derivation-3} and \eqref{Eq.Theorem2-Derivation-5} yields
\begin{equation} \label{Eq.Theorem2-Derivation-6}
\lim_{N \to \infty} \frac{\ln{P_{\mathrm{out}-\hat{\imath}}(t)}} {\ln\left(\frac{2c\beta\sigma_h^2}{\alpha-2}\left(\frac{r_{\rm{ex}}}{r_{\rm{net}}}\right)^2\,r_\mathrm{ex}^{-\alpha}\right) + \ln{N}} \to 1
\end{equation}
almost surely when $N \to \infty$. For completeness of exposition, the above result is summarized in the following theorem.

\begin{theorem}[Power scaling law for asymmetric UEs]
\label{Thoerem-heterogeneousCase}
When there are $N$ UEs uniformly distributed in the annulus of inner radius $r_\mathrm{ex}$ and outer radius $r_\mathrm{net}$, if the UE with the maximum instantaneous output DC power is  chosen to be charged during each transmission slot, then as $N$ becomes large asymptotically, the logarithm of the average output DC power scales according to
\begin{equation}
\label{ScalingLaw-heterogeneousCase}
\ln{\overline{P_{\mathrm{os}}^{\mathrm{het}}}}
\sim  \ln\left(\frac{2cN\beta\sigma_h^2}{\alpha-2}\left(\frac{r_{\rm{ex}}}{r_{\rm{net}}}\right)^2\,r_\mathrm{ex}^{-\alpha}\right).
\end{equation}
\end{theorem}

Next, by comparing \eqref{ScalingLaw-heterogeneousCase} with \eqref{AverageEnergy-Case2-Approx-1}, it is evident that
\begin{equation} \label{ScalingLaw-MobileCase-2}
\ln{\overline{P_{\mathrm{os}}^{\mathrm{het}}}}
\sim \ln\left(N\,\overline{P_{\mathrm{rrs}}^{\mathrm{het}}}\right).
\end{equation}
Afterwards, by using a similar approach to attain Corollary~\ref{Corolloary-fixedCase}, we have the following corollary.
\begin{corollary}[Power transfer efficiency for asymmetric UEs]
\label{Corolloary-MobileCase}
The opportunistic scheduling among $N$ asymmetric UEs yields a scheduling gain of~$N$ in terms of the power transfer efficiency, compared with the round-robin scheduling.
\end{corollary}

Comparing Corollary~\ref{Corolloary-MobileCase} with Corollary~\ref{Corolloary-fixedCase}, it is clear that the path-loss variety among asymmetric UEs leads to higher WPT efficiency. Specifically, if opportunistic scheduling is adopted, the power transfer efficiency increases from $\ln{N}$ in the homogeneous case (no path-loss variety among UEs) to $N$ in the heterogeneous scenario (with path-loss variety among UEs). Hence, by implementing opportunistic scheduling instead of the traditional round-robin scheduling at a power beacon, far-field power transfer efficiency can be significantly improved.

\begin{remark}[On the fairness among users in the heterogeneous scenario]
The coverage area of a power beacon is generally very small, compared with that of a BS. For instance, in the simulation setting detailed in Table~\ref{Table_ParameterSetting}, the radius of the coverage area of the considered power beacon is set to 30m and a typical UE is located 10m away from the beacon, whereas the radius of the BS cell is usually as large as 500m. Therefore, the power levels received from the beacon at different UEs are on the same order.  As a result, the fairness among heterogeneous users in the context of power transfer is almost guaranteed.
\end{remark}

\begin{table}[!t]
\caption{Parameter Setting Used in the Simulation Experiments}
\begin{center}
\begin{scriptsize}
\begin{tabular}{cccc}
\toprule
\textbf{Symbol}             & \textbf{Definition}                                                          & \textbf{Value}     & \textbf{Unit} \\
\midrule
$r_\mathrm{ex}$         &Radius of Exclusion Zone                                             &2                         &m                  \\
$r_\mathrm{net}$        &Radius of Circular Coverage                                         &30                       & m                 \\
$I_s$                            &Reverse Saturation Current of Schottky Diode            &1                          &mA                \\
$N$                             &Number of User Terminals                                             &1--50                   &                    \\
$P$                              &Transmit Power of the Power Beacon                          &43--53                  &dBm              \\
$V_T$                          &Thermal Voltage                                                            &28.85                    &mV                \\
$\alpha$                      &Path-Loss Exponent                                                       &4                          &                     \\
$\beta$                        &Shadowing Effect                                                            &1                            &                     \\
$\rho$                          &Quality Factor of Schottky Diode                                   &1                            &                     \\
$\xi$                             &Coefficient of Energy Conversion                                  &0.85                      &                     \\
$\sigma_h^2$             &Average Multi-Path Gain                                                 &1                            &                     \\
\bottomrule
\end{tabular}
\end{scriptsize}
\end{center}
\label{Table_ParameterSetting}
\end{table}

\section{Simulation Results and Discussions}
\label{Section:Simulation&Discussion}
In this section, we present and discuss simulation results, in comparison with numerical results pertaining to the previously developed analysis. The parameter setting used is summarized in Table~\ref{Table_ParameterSetting}. In particular, the radii of the exclusion zone and the circular coverage of the power beacon are set to $2$m and $30$m, respectively. Also, a typical user is set to be $10$m away from the power beacon, and its achieved power transfer efficiency is used to compare the effectiveness of different WPT schemes with or without scheduling. In the following, we first discuss simulation results obtained using round-robin scheduling.

\begin{remark}[Safety levels of human exposure to RF electromagnetic fields]
According to the IEEE Standard C95.1-2005, for safety levels with respect to human exposure to RF electromagnetic fields, the permissible exposure level from 2GHz to 100GHz in a public environment is 10$W/m^2$ \cite[p. 27]{IEEEStdC95.1}. Typical frequencies that are used for prototype development of far-field WPT systems are 2.45GHz and 5.8GHz (because they belong to the ISM bands), fall exactly into the aforementioned frequency range. In our simulation setting (Table~\ref{Table_ParameterSetting}), if the effective radiated power ($P_\mathrm{erp}$) at the power beacon takes the maximum value 53dBm, i.e. 200W, the power densities at the distances from the power beacon $d=2m$ (the closest allowable), $10m$ (distance of a typical UE) and $30m$ (edge of beacon coverage), are 3.98$W/m^2$, 0.159$W/m^2$ and 0.0177$W/m^2$, respectively, by recalling that power density is computed via $P_D = P_\mathrm{erp}/(4\pi d^2)$. Evidently, these densities are smaller than 10$W/m^2$ and, thus, permissible in practice.
\end{remark}

\subsection{Round-Robin Scheduling}
\label{Sim_RoundRobin}

\begin{figure}[t]
\centering
\includegraphics [width=3.75in, clip, keepaspectratio]{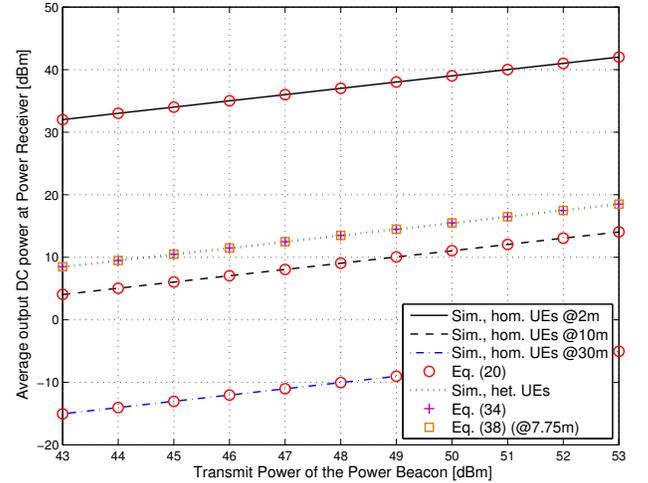}
\caption{The average output DC power when round-robin scheduling is performed among homogeneous (`hom') or heterogeneous (`het') UEs.}
\label{Fig_RoundRobin}
\end{figure}

Figure~\ref{Fig_RoundRobin} illustrates the average output DC power when round-robin scheduling is applied among multiple symmetric or asymmetric UEs, as a function of the beacon transmit power. The transmit power (X-axis) and the average output DC received power (Y-axis) are shown in dBm, i.e. $P = 10\log_{10}(P_0/10^{-3})$ where $P_0$ is in the unit of watt. For illustration purposes, the distances of the symmetric UEs are set to 2m (i.e. at the inner boundary of the annular coverage as shown in Fig.~\ref{Fig_Scheduling}), 10m and 30m (i.e. at the outer boundary of the annular coverage). It is observed  from Fig.~\ref{Fig_RoundRobin} that the average output DC power decreases sharply with the increase in distance and that the simulation results are in perfect agreement with the numerical results computed by using \eqref{AverageEnergy-Case1}.

On the other hand, as shown in the dashed plot and with the plus/square marks, the heterogeneous scenario harvests the same average power as the symmetric UEs located at $7.75$m (computed by \eqref{DistanceScalingFator}) away from the power beacon, which coincides with the numerical results computed by \eqref{AverageEnergy-Case2}. Consequently, for any particular coverage with inner and outer boundaries, the analytical expression shown in \eqref{AverageEnergy-Case1} (or \eqref{AverageEnergy-Case2}) can be applied to efficiently predict the average output DC power in homogeneous (or heterogeneous) scenario, if round-robin scheduling is applied. Also, the effect of channel variety among asymmetric UEs can be exactly described by the distance scaling factor shown in \eqref{DistanceScalingFator}.

\begin{figure}[t]
\centering
\includegraphics [width=3.75in, clip, keepaspectratio]{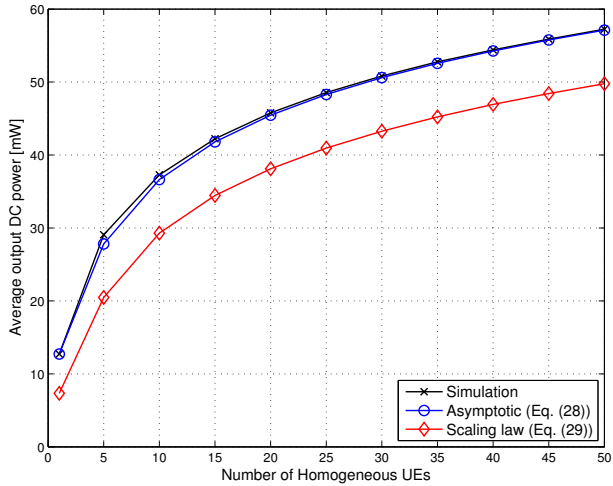}
\caption{The average output DC power (in the unit of mW) when opportunistic scheduling is performed among homogeneous UEs ($d=10\rm{m}$), with the transmit power of the power beacon being 50dBm.}
\label{Fig_OpportunisticSymmetric}
\end{figure}

\subsection{Opportunistic Scheduling}
\label{Sim_Opportunistic}

\begin{figure}[t]
\centering
\includegraphics [width=3.75in, clip, keepaspectratio]{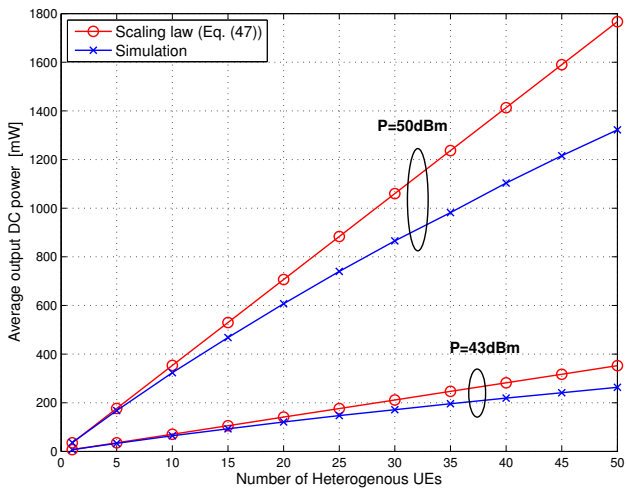}
\caption{The average output DC power when opportunistic scheduling is performed among heterogeneous UEs, in the unit of mW.}
\label{Fig_OpportunisticAsymmetric}
\end{figure}

With opportunistic scheduling performed among symmetric UEs, Fig.~\ref{Fig_OpportunisticSymmetric} shows the average output DC power in linear scale, i.e. in the unit of mW ($1\rm{mW} = 10^{-3} \rm{watt}$), where the beacon transmit power is set to $50$dBm and the clustered UEs are located $10$m away from the power beacon. It is clear that the simulation results of the average DC power agree exactly with the numerical results computed by \eqref{AverageEnergy-Case3}. On the other hand, the scaling law given by \eqref{ScalingLaw-homogeneousCase} reflects the increasing trend of the DC power very well and, more importantly, all three curves demonstrate that the average DC power increases {\it logarithmically} with the number of UEs, as predicted by Theorem~\ref{Thoerem-homogeneousCase}. Finally, we note that the small gap between the simulation and the numerical results of the scaling law, as shown in Fig.~\ref{Fig_OpportunisticSymmetric}, is due to ignoring the Euler's constant in \eqref{AverageEnergy-Case3} when \eqref{ScalingLaw-homogeneousCase} was derived. Also, for the special case of single UE, i.e. when $N=1$ (corresponding to the most-left point of the lowest curve in Fig.~\ref{Fig_OpportunisticSymmetric}), the numerical result of the scaling law  was computed as per \eqref{AverageEnergy-Case3}, since, in such a special case, the scaling law shown in \eqref{ScalingLaw-homogeneousCase} degenerates to zero.

On the other hand, if opportunistic scheduling is performed among multiple asymmetric UEs, Fig.~\ref{Fig_OpportunisticAsymmetric} illustrates the average output DC power in the unit of mW, with respect to the number of UEs $N$. It is observed that, for either values of the beacon transmit power, 43dBm or 50dBm, the power scaling law computed by \eqref{ScalingLaw-heterogeneousCase} perfectly reflects the increasing trend of the simulation results, i.e. {\it linearly} increasing with $N$. Moreover, when the transmit power is increased from 43dBm to 50dBm, if we show the output DC power in dBm, then it is easy to find that the average output power corresponding to the case with the beacon transmit power set to $P=50$dBm is exactly $7$dB larger than that of the case with $P=43$dBm, i.e. the same amount of increase as in the transmit power of the beacon. This is not strange since the output DC power at a particular receiver always increases linearly with the beacon transmit power, as shown in \eqref{HarvestedEnergy}.

Compared with Fig.~\ref{Fig_OpportunisticSymmetric}, it is evident that the curves in Fig.~\ref{Fig_OpportunisticAsymmetric} have steeper slopes. This is because opportunistic scheduling yields larger diversity gain than round-robin scheduling, namely $N$ versus $\ln{N}$, as stated in Corollaries 1 and~2.

\begin{figure}[t]
\centering
\includegraphics [width=3.75in, clip, keepaspectratio]{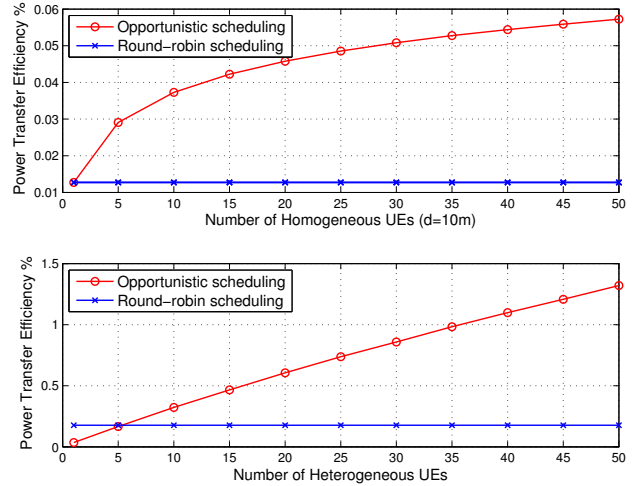}
\caption{The power transfer efficiency.}
\label{Fig_PowerEfficiency}
\end{figure}

\subsection{Power Transfer Efficiency}
Based on the results shown in Figs.~\ref{Fig_OpportunisticSymmetric} and \ref{Fig_OpportunisticAsymmetric}, the power transfer efficiency when opportunistic scheduling is applied among symmetric UEs is shown in the upper panel of Fig.~\ref{Fig_PowerEfficiency} and that among asymmetric UEs is displayed in the lower panel, in comparison with round-robin scheduling. The power transfer efficiencies pertaining to the round-robin mode are computed by $\overline{P_{\mathrm{rrs}}^{\mathrm{hom}}}/P_{\mathrm{erp}}$, where $\overline{P_{\mathrm{rrs}}^{\mathrm{hom}}}$ is given by \eqref{AverageEnergy-Case1} in the symmetric case and by \eqref{AverageEnergy-Case2} in the asymmetric one.

From the upper panel, it is observed that, the WPT efficiency with opportunistic scheduling increases logarithmically with the number of UEs while that of round-robin scheduling is constant at about $0.013\%$. On the other hand, the curves in the lower panel show that the WPT efficiency with opportunistic scheduling increases linearly with the number of UEs while that of round-robin scheduling remains about $0.18\%$. In particular, when the number of UEs $N>35$, the WPT efficiency under opportunistic scheduling is higher than the $1\%$ value which serves as a benchmark to the success of far-field WPT. Therefore, opportunistic scheduling is demonstrated to be a powerful technique to significantly improve far-field WPT efficiency. Finally, it is worthwhile to point out that the appearance of the WPT efficiency under round-robin scheduling looking higher  than that under opportunistic scheduling when $N$ is smaller than 5 (cf. lower panel of Fig.~\ref{Fig_PowerEfficiency}) is due to the fact that the average output DC power shown in \eqref{ScalingLaw-heterogeneousCase} is an asymptotic result, which is not accurate when $N$ is small.

\section{Conclusion}
\label{Section:Conclusion}
This paper provided a comprehensive study on the transmission efficiency of far-field wireless power transfer (WPT), by taking into account both large-scale path loss and small-scale multi-path fading over bounded coverage area. The analytically computed time-average output DC powers with respect to round-robin scheduling and opportunistic scheduling demonstrate that the latter scheduling strategy can improve far-field WPT efficiency linearly with the number of user equipments. This result provides a new approach to significantly improve the end-to-end power transfer efficiency of far-field WPT systems, apart from the traditional approach of increasing the transmit and receive antenna gains.

\appendices
\section{Proof of Lemma~\ref{Lemma1}}
\label{AppendixA}
Before detailing the proof of Lemma~\ref{Lemma1} in Section~\ref{ScenarioA-Opportunistic}, we recall one of the seminal von Mises criteria \cite{Falk93}, which will be used later in the form of Lemma~\ref{LemmaA1}.

\setcounter{lemma}{0}
\renewcommand{\thelemma}{A\arabic{lemma}}

\begin{lemma}[von Mises sufficient conditions for the domain of attraction of Gumbel distribution]
\label{LemmaA1}
Let $F(x)$ be a distribution function, if there is a real number $x_1$ such that, for all $x_1 \le x < \omega(F)$ where $\omega(F) = \mathrm{sup}\{x: F(x) < 1\}$, $f(x) = F^\prime(x) > 0$ , and
\begin{equation} \label{Eq.AppendixB-1}
\lim_{x \to \omega(F)}\frac{f(x)}{1-F(x)} = c_0,
\end{equation}
where $c_0 \in (0, \infty)$ is a constant, then, $F(x)$ is in the domain of attraction of the Gumbel distribution function $H_{3, \, 0}(x) \triangleq \exp\left(-e^{-x}\right)$.
\end{lemma}

Applying Lemma~\ref{LemmaA1} to the CDF $F_{P_{\mathrm{out}-i}(t)}(x)$ \eqref{CDF-Energy-Case1}, it is clear that the parameters used in Lemma~\ref{LemmaA1} can be given by $x_1 = 0$ and $\omega(F) = \infty$. Then, when $0 \le x < \infty$, the first-order derivative of $F_{P_{\mathrm{out}-i}(t)}(x)$ exists, i.e. $f_{P_{\mathrm{out}-i}(t)}(x)$ in \eqref{PDF-Energy-Case1}, and is greater than zero. Hence, the first condition of the von Mises criterion is satisfied. Further, by virtue of the PDF in \eqref{PDF-Energy-Case1} and the CDF in \eqref{CDF-Energy-Case1}, it is easy to show that
\begin{equation} \label{Eq.AppendixB-2}
\lim_{x \to \infty}\frac{f_{P_{\mathrm{out}-i}(t)}(x)}{1-F_{P_{\mathrm{out}-i}(t)}(x)}
= \frac{d^\alpha}{c\beta\sigma_h^2},
\end{equation}
which implies that the second condition of the von Mises criterion is also satisfied. Consequently, by recalling Lemma~\ref{LemmaA1}, $F_{P_{\mathrm{out}-i}(t)}(x)$ given by \eqref{CDF-Energy-Case1} is in the domain of attraction of the Gumbel distribution function $H_{3, \, 0}(x)$. That is, the limiting distribution function of the maximum among $N$ output DC powers, i.e. $P_{\mathrm{out}-\hat{\imath}}(t) = \max\{P_{\mathrm{out}-1}(t), P_{\mathrm{out}-2}(t), \cdots, P_{\mathrm{out}-N}(t)\}$, is given by the Gumbel distribution as shown in \eqref{Eq.LimitingCase3}. Furthermore, the scaling factor $a_1$ is determined by \cite[Theorem 2.1.3]{Galambos87}
\begin{equation} \label{Eq.AppendixB-3}
a_1 = \inf\left\{x: 1-F_{P_{\mathrm{out}-i}(t)}(x) \le \frac{1}{N}\right\}.
\end{equation}
Since the CDF given by \eqref{CDF-Energy-Case1} is monotonically increasing, substituting \eqref{CDF-Energy-Case1} into \eqref{Eq.AppendixB-3}  and performing some algebraic manipulations yields \eqref{Eq.LimitingCase3-a}. The positioning parameter $b_1$ is determined by \cite[pp.~104--105]{Galambos87}
\begin{equation} \label{Eq.AppendixB-4}
b_1 = \frac{1-F_{P_{\mathrm{out}-i}(t)}(a)}{f_{P_{\mathrm{out}-i}(t)}(a)}.
\end{equation}
Substituting \eqref{PDF-Energy-Case1}--\eqref{CDF-Energy-Case1} into \eqref{Eq.AppendixB-4} results in \eqref{Eq.LimitingCase3-b}, thus completing the proof.

\section{Derivation of Eq.~\eqref{CDF-Energy-Case2}}
\label{Appendix-B}
We first derive the CDF and the PDF of $d_{i}^{-\alpha}(t)$. By definition, the CDF of $d_{i}^{-\alpha}(t)$ can be given by
\begin{equation}  \label{Appendix-A-1}
F_{d_{i}^{-\alpha}(t)}(x)
= \mathrm{Pr}\left\{d_{i}^{-\alpha}(t) < x\right\} 	
= 1- F_{d_{i}(t)}\left(x^{-\frac{1}{\alpha}}\right),			
\end{equation}
where $r_\mathrm{ex} \le x^{-\frac{1}{\alpha}} \le r_\mathrm{net}$, i.e. $r_\mathrm{net}^{-\alpha}\le x \le r_\mathrm{ex}^{-\alpha}$. Substituting \eqref{CDF-Distance} into \eqref{Appendix-A-1} and performing some algebraic manipulations, we obtain
\begin{equation} \label{CDF-Distance-2}
F_{d_{i}^{-\alpha}(t)}(x)
= \begin{cases}
\frac{r_\mathrm{net}^2 - x^{-\frac{2}{\alpha}}}{r_\mathrm{net}^2 - r_\mathrm{ex}^2},	& \text{if } r_\mathrm{net}^{-\alpha}\le x \le r_\mathrm{ex}^{-\alpha}, \\
0,	& \text{otherwise}.
\end{cases}
\end{equation}
Then, differentiating \eqref{CDF-Distance-2} with respect to $x$ yields the PDF:
\begin{equation} \label{PDF-Distance-2}
f_{d_{i}^{-\alpha}(t)}(x)
= \begin{cases}
\frac{2}{\alpha(r_\mathrm{net}^2 - r_\mathrm{ex}^2)}\,x^{-\frac{2}{\alpha}-1},	& \text{if } r_\mathrm{net}^{-\alpha}\le x \le r_\mathrm{ex}^{-\alpha}, \\
0,	& \text{otherwise}.
\end{cases}
\end{equation}

Now, we derive the CDF of the output DC power $P_{\mathrm{out}-i}(t) = c\beta{d_{i}^{-\alpha}(t)}|h_{i}(t)|^2$.  By definition, we have
\begin{small}
\begin{eqnarray}
\lefteqn{F_{P_{\mathrm{out}-i}(t)}(x)}     \nonumber \\
&  =  & \mathrm{Pr}\left\{c\beta{d_{i}^{-\alpha}(t)}|h_{i}(t)|^2 < x\right\} 				\nonumber \\
&  =  & \int\limits_0^\infty{\mathrm{Pr}\left\{c\beta{d_{i}^{-\alpha}(t)}|h_{i}(t)|^2 < x \vert |h_{i}(t)|^2=y\right\} f_{|h_{i}(t)|^2}(y)}\,\mathrm{d}y		\nonumber \\
&  =  & \int\limits_0^\infty{\mathrm{Pr}\left\{{d_{i}^{-\alpha}(t)} < \frac{x}{c\beta y} \right\} f_{|h_{i}(t)|^2}(y)}\,\mathrm{d}y		\nonumber \\
&  =  & \int\limits_0^\infty{F_{d_{i}^{-\alpha}(t)}\left(\frac{x}{c\beta y} \right) f_{|h_{i}(t)|^2}(y)}\,\mathrm{d}y		 \nonumber \\
&  =  & \int\limits_0^{\frac{x}{c\beta}r^\alpha_\mathrm{ex}}{\hspace{-10pt} f_{|h_{i}(t)|^2}(y)}\,\mathrm{d}y
+ \int\limits_{\frac{x}{c\beta}r^\alpha_\mathrm{ex}}^{\frac{x}{c\beta}r^\alpha_\mathrm{net}}
{\hspace{-10pt} F_{d_{i}^{-\alpha}(t)}\left(\frac{x}{c\beta y} \right) f_{|h_{i}(t)|^2}(y)}\,\mathrm{d}y		 \label{Appendix-B-2} \\
&  =  & 1-\exp\left(-\frac{r_\mathrm{ex}^\alpha}{c \beta\sigma_h^2}\,x\right)
+ \underbrace{\int_{\frac{x}{c\beta}r^\alpha_\mathrm{ex}}^{\frac{x}{c\beta}r^\alpha_\mathrm{net}}
{\frac{r_\mathrm{net}^2}{r_\mathrm{net}^2 - r_\mathrm{ex}^2} f_{|h_{i}(t)|^2}(y)}\,\mathrm{d}y}_{I_1}		 \nonumber \\
&     &{} - \underbrace{\int_{\frac{x}{c\beta}r^\alpha_\mathrm{ex}}^{\frac{x}{c\beta}r^\alpha_\mathrm{net}}
{\frac{1}{r_\mathrm{net}^2 - r_\mathrm{ex}^2} \left(\frac{x}{c\beta y} \right)^{-\frac{2}{\alpha}} f_{|h_{i}(t)|^2}(y)}\,\mathrm{d}y}_{I_2} \, ,	\label{Appendix-B-3}
\end{eqnarray}
\end{small}
\hspace{-6pt} where \eqref{CDF-Distance-2} was inserted into \eqref{Appendix-B-2} to reach \eqref{Appendix-B-3}. Subsequently, in light of \eqref{PDF-Nakagami}, the first integral term in \eqref{Appendix-B-3} is computed as
\begin{eqnarray}	
I_1
&  =  & \frac{r_\mathrm{net}^2}{r_\mathrm{net}^2 - r_\mathrm{ex}^2} \int_{\frac{x}{c\beta}r^\alpha_\mathrm{ex}}^{\frac{x}{c\beta}r^\alpha_\mathrm{net}}     \nonumber
{\frac{1}{\sigma_h^2} \exp\left(-\frac{y}{\sigma_h^2}\right)}\,\mathrm{d}y		\label{Appendix-B-4} \\
&  =  & \frac{r_\mathrm{net}^2}{r_\mathrm{net}^2 - r_\mathrm{ex}^2}
\left[\exp\left(-\frac{r^\alpha_\mathrm{ex}}{c\beta\sigma_h^2}\,x\right) \right. \nonumber \\
&      & \left. - \exp\left(-\frac{r^\alpha_\mathrm{net}}{c\beta\sigma_h^2}\,x\right)\right].		 \label{Appendix-B-5}
\end{eqnarray}
Also, the second integral term in \eqref{Appendix-B-3} can be computed as
\begin{eqnarray}	
I_2
&  =  & \frac{1}{r_\mathrm{net}^2 - r_\mathrm{ex}^2} \int_{\frac{x}{c\beta}r^\alpha_\mathrm{ex}}^{\frac{x}{c\beta}r^\alpha_\mathrm{net}}
{\left(\frac{x}{c\beta y} \right)^{-\frac{2}{\alpha}} \frac{1}{\sigma_h^2} \exp\left(-\frac{y}{\sigma_h^2}\right)}\,\mathrm{d}y		\nonumber \\
&  =  & \frac{1}{r_\mathrm{net}^2 - r_\mathrm{ex}^2} \left(\frac{x}{c\beta\sigma_h^2} \right)^{-\frac{2}{\alpha}}\left[\Gamma\left(1+\frac{2}{\alpha}, \, \frac{r^\alpha_\mathrm{ex}}{c\beta\sigma_h^2}\,x\right)\right.  \nonumber \\
&      &{} - \left.\Gamma\left(1+\frac{2}{\alpha}, \, \frac{r^\alpha_\mathrm{net}}{c\beta\sigma_h^2}\,x\right)\right].		 \label{Appendix-B-6}
\end{eqnarray}
Finally, inserting \eqref{Appendix-B-5}--\eqref{Appendix-B-6} into \eqref{Appendix-B-3} and performing some algebraic manipulations yields the desired \eqref{CDF-Energy-Case2}.

\section{Proof of Lemma~\ref{Lemma2}}
\label{AppendixC}
Before detailing the proof of Lemma~\ref{Lemma2}, we reproduce another von Mises criterion different from Lemma~\ref{LemmaA1}  in Appendix~\ref{AppendixA} as the following Lemma~\ref{LemmaC1} \cite[Theorem 10.5.2]{David03}.

\setcounter{lemma}{0}
\renewcommand{\thelemma}{C\arabic{lemma}}

\begin{lemma}[von Mises sufficient conditions for the domain of attraction of Fr\'{e}chet distribution]
\label{LemmaC1}
Let $F(x)$ be a distribution function with $\omega(F) = \infty$, if $f(x) = F^\prime(x) > 0$ for all large $x$ and for some $\zeta \in (0, \infty)$,
\begin{equation} \label{Eq.AppendixC-1}
\lim_{x \to \infty}\frac{x f(x)}{1-F(x)} = \zeta,
\end{equation}
then, $F(x)$ is in the domain of attraction of the Fr\'{e}chet function $H_{1, \, \zeta}(x) \triangleq \exp\left(-x^{-\zeta}\right),~x > 0$.
\end{lemma}

Recalling the fact that the first-order derivative of the complementary incomplete Gamma function is given by $\frac{\mathrm{d}}{\mathrm{d}\,x}\, \Gamma(a, bx) = -b\, e^{-b x} \,(b x)^{-1 + a}$ and taking the derivative of \eqref{CDF-Energy-Case2} with respect to $x$, yields the PDF of the output DC power. Then, substituting \eqref{CDF-Energy-Case2} and its PDF into \eqref{Eq.AppendixC-1}, and performing some algebraic manipulations, it is easy to show that $\zeta = \frac{2}{\alpha}$. Therefore, the limiting distribution function of the maximum among $N$ output DC powers, i.e. $P_{\mathrm{out}-\hat{\imath}}(t) = \max\{P_{\mathrm{out}-1}(t), P_{\mathrm{out}-2}(t), \cdots, P_{\mathrm{out}-N}(t)\}$, is given by the Fr\'{e}chet distribution as shown in \eqref{Eq.LimitingCase4}. Furthermore, the scaling factor $b_2$ in \eqref{Eq.LimitingCase4} is determined by \cite[Eq.~(10.5.6)]{David03}
\begin{equation} \label{Eq.AppendixC-2}
b_2 = \inf\left\{x: 1-F_{P_{\mathrm{out}-i}(t)}(x) \le \frac{1}{N}\right\}.
\end{equation}
Due to the high complexity of $F_{P_{\mathrm{out}-i}(t)}(x)$ shown in \eqref{CDF-Energy-Case2}, the exact solution to $b_2$ is mathematically intractable. However, an explicit approximation of $b_2$ can be derived as follows.

In view of \eqref{CDF-Energy-Case2}, we have \eqref{Eq.AppendixC-4} at the top of the next page,
\begin{eqnarray}	
\lefteqn{1-F_{P_{\mathrm{out}-i}(t)}(x)}  \nonumber \\
&  =  & \frac{1}{r_\mathrm{net}^2 - r_\mathrm{ex}^2} \left[r_\mathrm{net}^2 \exp\left(-r_2\,x\right) - r_\mathrm{ex}^2 \exp\left(-r_1\,x\right)\right]		\nonumber \\
&      & {}+\frac{1}{r_\mathrm{net}^2 - r_\mathrm{ex}^2}\left(\frac{c\beta\sigma_h^2}{x} \right)^{\frac{2}{\alpha}}   \nonumber \\
&      &{}\times
\left[\Gamma\left(1+\frac{2}{\alpha}, \, r_1\,x\right)
 - \Gamma\left(1+\frac{2}{\alpha}, \, r_2\,x\right)\right]			\nonumber \\
&  \approx & \frac{2\,r_\mathrm{ex}^2}{\alpha(r_\mathrm{net}^2 - r_\mathrm{ex}^2)}\frac{\exp(-r_1\,x)}{r_1\,x}  \nonumber \\
&      & {}- \frac{2\,r_\mathrm{net}^2}{\alpha(r_\mathrm{net}^2 - r_\mathrm{ex}^2)}\frac{\exp(-r_2\,x)}{r_2\,x},		 \label{Eq.AppendixC-4}
\end{eqnarray}
where the asymptotic series of the incomplete Gamma function, i.e. $\Gamma(a, x) = e^{-x}\,x^{a-1}\left[1+\frac{a-1}{x}+{\it O}\left(\frac{1}{x^2}\right)\right]$, was employed to get \eqref{Eq.AppendixC-4}. Moreover, since $r_2 \gg r_1$ as defined after \eqref{CDF-Energy-Case2}, the second term on the right-hand side of \eqref{Eq.AppendixC-4} can be further ignored (this will be compensated later), yielding
\begin{eqnarray}	
1-F_{P_{\mathrm{out}-i}(t)}(x)
& \approx  & \frac{2\,r_\mathrm{ex}^2}{\alpha(r_\mathrm{net}^2 - r_\mathrm{ex}^2)}\frac{\exp(-r_1\,x)}{r_1\,x} \nonumber \\
& \approx  & \frac{2}{\alpha} \left(\frac{r_\mathrm{ex}}{r_\mathrm{net}}\right)^2\frac{\exp(-r_1\,x)}{r_1\,x}. \label{Eq.AppendixC-5}
\end{eqnarray}

Then, by recalling the fact that the CDF $F_{P_{\mathrm{out}-i}(t)}(x)$ is monotonically increasing with respect to $x$,  substituting \eqref{Eq.AppendixC-5} into \eqref{Eq.AppendixC-2} and performing some algebraic manipulations yield
\begin{equation}	\label{Eq.AppendixC-6}
r_1b_2\exp(r_1b_2)
\approx \frac{2N}{\alpha} \left(\frac{r_\mathrm{ex}}{r_\mathrm{net}}\right)^2.
\end{equation}
Since $\frac{2N}{\alpha} \left(\frac{r_\mathrm{ex}}{r_\mathrm{net}}\right)^2 < \frac{1}{e}$, with $e = 2.7183 \cdots$ being the Euler's number, holds in practice, then applying the series expansion of the Lambert function \cite[Eq. (4.13.5)]{Olver10} to \eqref{Eq.AppendixC-6} results in
\begin{equation}	\label{Eq.AppendixC-7}
b_2
\approx \frac{2N}{r_1\,\alpha} \left(\frac{r_\mathrm{ex}}{r_\mathrm{net}}\right)^2
= \frac{2cN\beta\sigma_h^2}{\alpha}\left(\frac{r_\mathrm{ex}}{r_\mathrm{net}}\right)^2 r_\mathrm{ex}^{-\alpha}.
\end{equation}

Notice that ignoring the second term in \eqref{Eq.AppendixC-4} results in the value of $b_2$ being underestimated, since the function $\exp(-x)/x$ involved in the first term of \eqref{Eq.AppendixC-4} is monotonically decreasing with $x > 0$. To compensate this approximation error, $b_2$ in \eqref{Eq.AppendixC-7} is empirically enlarged (based on the observation from \eqref{AverageEnergy-Case2-Approx-1}) so as to obtain the desired \eqref{Eq.LimitingCase4-b}. The effectiveness of $b_2$ in \eqref{Eq.LimitingCase4-b} is corroborated by ensuing simulation results in Section \ref{Sim_Opportunistic}. Finally, it is noted that the value of $b_2$ is not unique and that different choices lead to different convergence speeds of the limiting distribution function \cite[Section 2.10]{Galambos87}.

\bibliographystyle{IEEEtran}
\bibliography{References}

\begin{IEEEbiography}[{\includegraphics[width=1in, height=1.25in,  clip, keepaspectratio]{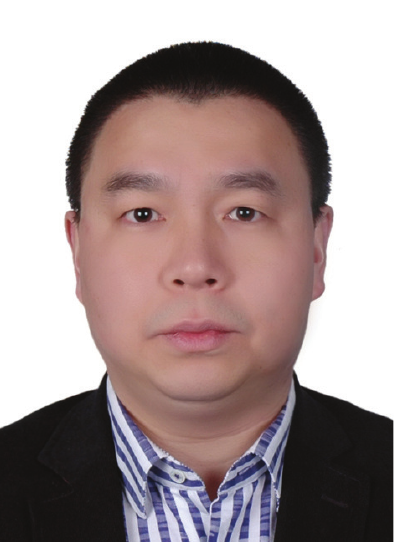}}]{Minghua Xia} (M'12)
obtained his Ph.D. degree in Telecommunications and Information Systems from SUN Yat-sen University, Guangzhou, China, in 2007. Since 2015, he has been working as a Professor at the same university.

From 2007 to 2009, he was with the Electronics and Telecommunications Research Institute (ETRI) of South Korea, Beijing R\&D Center, Beijing, China, where he worked as a member and then as a senior member of engineering staff and participated in the projects on the physical layer design of 3GPP LTE mobile communications. From 2010 to 2014, he was in sequence with The University of Hong Kong, Hong Kong, China; King Abdullah University of Science and Technology, Jeddah, Saudi Arabia; and the Institut National de la Recherche Scientifique (INRS-EMT), University of Quebec, Montreal, Canada, as a Postdoctoral Fellow. His research interests are in the general area of 5G wireless communications, and in particular the design and performance analysis of multi-antenna systems, cooperative relaying systems and cognitive relaying networks, and recently focus on the design and analysis of wireless power transfer and/or energy harvesting systems, as well as massive MIMO and small cells. He holds two patents granted in China.

Dr. Xia was awarded as Exemplary Reviewer by IEEE Transactions on Communications, IEEE Communications Letters, and IEEE Wireless Communications Letters, respectively, in 2014.
\end{IEEEbiography}

\vfill

\begin{IEEEbiography}
[{\includegraphics[width=1in, height=1.25in,  clip, keepaspectratio]{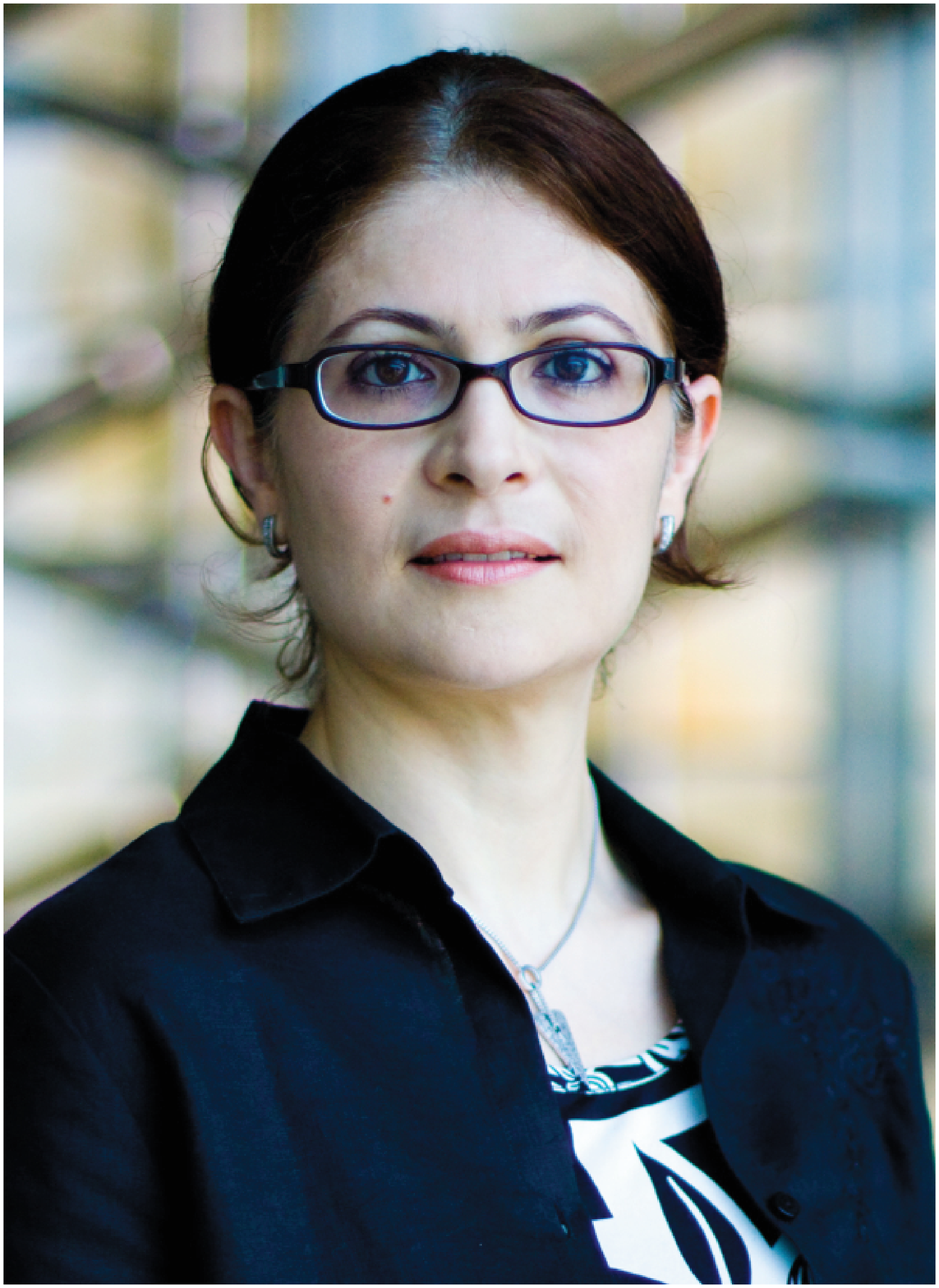}}]{Sonia A\"{\i}ssa} (S'93-M'00-SM'03) received her Ph.D. degree in Electrical and Computer Engineering from McGill University, Montreal, QC, Canada, in 1998. Since then, she has been with the Institut National de la Recherche Scientifique-{\it Energy, Materials and Telecommunications} Center (INRS-EMT), University of Quebec, Montreal, QC, Canada, where she is a Full Professor.

From 1996 to 1997, she was a Researcher with the Department of Electronics and Communications of Kyoto University,
and with the Wireless Systems Laboratories of NTT, Japan. From 1998 to 2000, she was a Research Associate at INRS-EMT. In 2000-2002, while she was an Assistant Professor, she was a Principal Investigator in the major program of personal and mobile communications of the Canadian Institute for Telecommunications Research, leading research in radio resource management for wireless networks. From 2004 to 2007, she was an Adjunct Professor with Concordia University, Montreal. In 2006, she was Visiting Invited Professor at the Graduate School of Informatics, Kyoto University, Japan. Her research interests include the modeling, design and performance analysis of wireless communication systems and networks.

Dr. A\"{\i}ssa is the Founding Chair of the IEEE Women in Engineering Affinity Group in Montreal,  2004-2007; acted as TPC Leading Chair or Cochair of the Wireless Communications Symposium at IEEE ICC '06, '09, '11, '12; PHY/MAC Program Cochair of the 2007 IEEE WCNC; TPC Cochair of the 2013 IEEE VTC-spring; and TPC Symposia Chair of the 2014 IEEE Globecom. Her main editorial activities include: Editor, {\scshape IEEE Transactions on Wireless Communications}, 2004-2012; Associate Editor, {\scshape IEEE Communications Magazine}, 2004-2009; Technical Editor, {\scshape IEEE Wireless Communications Magazine}, 2006-2010; and Associate Editor, {\it Wiley Security and Communication Networks Journal}, 2007-2012. She currently serves as Area Editor for the {\scshape IEEE Transactions on Wireless Communications}, and Technical Editor for the {\scshape IEEE Communications Magazine}. Awards to her credit include the NSERC University Faculty Award in 1999; the Quebec Government FQRNT Strategic Faculty Fellowship in 2001-2006; the INRS-EMT Performance Award multiple times since 2004, for outstanding achievements in research, teaching and service; and the Technical Community Service Award from the FQRNT Centre for Advanced Systems and Technologies in Communications in 2007. She is co-recipient of five IEEE Best Paper Awards and of the 2012 IEICE Best Paper Award; and recipient of NSERC Discovery Accelerator Supplement Award. She is a Distinguished Lecturer of the IEEE Communications Society (ComSoc) and an Elected Member of the ComSoc Board of Governors.
\end{IEEEbiography}

\vfill

\end{document}